\documentclass[11pt]{article}

\usepackage{tikz}
\usepackage{authblk}
\usepackage{geometry}
\usepackage[english]{babel}
\usepackage[utf8]{inputenc}
\usepackage[T1]{fontenc}
\usepackage{indentfirst}
\usepackage{amsmath}
\usepackage{amssymb}
\usepackage{amsthm}
\usepackage{proof}
\usepackage{eufrak}
\usepackage[font=small,labelfont=bf]{caption}

\allowhyphens

\newcommand{\complex}{\mathbb{C}}
\newcommand{\egesz}{\mathbb{Z}}

\newcommand{\valos}{\mathbb{R}}

\renewcommand{\ge}{\geqslant}
\renewcommand{\le}{\leqslant}

\newcommand{\ket}[1]{{\left|#1\right\rangle}}
\newcommand{\bra}[1]{{\left\langle #1\right|}}

\newcommand{\skalarszorzat}[2]{{\langle #1 | #2 \rangle}}

\newcommand{\oo}{o}

\newcommand{\ZZ}{\mathcal{Z}}
\newcommand{\Sc}{\mathcal{S}}
\newcommand{\HH}{\mathcal{H}}
\newcommand{\PP}{\mathcal{P}}


\usepackage{ifpdf}

\ifpdf
\usepackage{epstopdf}
\usepackage[pdftex,colorlinks,urlcolor=blue,citecolor=blue,linkcolor=blue]{hyperref}
\else
\usepackage[hypertex,colorlinks,urlcolor=blue,citecolor=blue,linkcolor=blue]{hyperref}
\fi
\begin{document}
\numberwithin{equation}{section}

\title{
    Multi-directional unitarity and maximal entanglement in spatially symmetric quantum states  
  }
\author[1]{M\'arton Mesty\'an}
\author[1]{Bal\'azs Pozsgay}
\author[2]{Ian M. Wanless}

\affil[1]{MTA-ELTE “Momentum” Integrable Quantum Dynamics Research Group, Department of Theoretical Physics, Eötvös
  Loránd University, Hungary}
\affil[2]{School of Mathematics, Monash University, Australia
}

\maketitle

\abstract{
  We consider dual unitary operators and their multi-leg generalizations that have appeared at various places in the
  literature. These objects can be related to multi-party quantum states with special entanglement patterns: the sites are
  arranged in a spatially symmetric pattern and the states have maximal entanglement for all bipartitions that follow
  from the reflection symmetries of the given geometry.
  We consider those cases where the state itself is
invariant with respect to the geometrical symmetry group. The simplest examples are those dual unitary operators which
are also self dual and reflection invariant, but we also consider the generalizations in the hexagonal,
cubic, and octahedral geometries. 
We provide a number of constructions and concrete examples for these
objects for various local dimensions. All of our examples can be used to build quantum
cellular automata in 1+1 or 2+1 
dimensions, with multiple equivalent choices for the ``direction of time''. 
}

\section{Introduction}

In this work we treat certain quantum mechanical objects with
very special entanglement
properties. { We call these objects} ``multi-directional unitary operators'', or alternatively
``multi-directional maximally entangled states''. 
{These objects have relevance in both quantum many body physics and quantum information theory, and they have appeared in multiple works in the literature of both fields.}
{On one hand,} the {objects} that  we treat are straightforward generalizations of the
well-studied ``dual unitary'' (or bi-unitary) operators which appeared in the study of solvable many body systems,  in
particular solvable quantum cellular automata 
\cite{dual-unitary-1,dual-unitary-2,dual-unitary-3}. On the other hand, these objects are also generalizations of the
``absolutely maximally entangled states'' studied in quantum information theory \cite{AME-review}.
{Beyond the above two fields, the related concept of bi-unitarity also appeared much earlier} in the study of von Neumann
algebras \cite{jones-planar}. In this introduction we first present the different approaches to the objects of our
study, and afterwards we specify 
the goals of this work. 

{In the field of solvable quantum many body systems, our study generalizes the concept of ``dual unitarity''.} Dual unitary quantum circuits are solvable one dimensional many body systems which exist in discrete space and discrete
time. They are local (Floquet) quantum circuits {made up from the repetitions of the same} two-site {quantum gate, which is unitary not only in the time, but \emph{also in the space direction.} Such circuits describe a unitary evolution in both the time and the space dimensions. This} concept of dual unitarity was formulated in
\cite{dual-unitary-3}, and it is based on earlier work on a specific model, the kicked Ising model (see
\cite{dual-unitary-1,dual-unitary-2} and also
\cite{kicked-Ising-space-time-duality-1,sarang-lamacraft-kicked-DU}). Besides the possibility of computing exact
correlation functions \cite{dual-unitary-3} (including non-equilibrium situations \cite{dual-unitary-4}) the dual
unitary models give access also to the entanglement evolution for states and also for local operators; generally, these
models show maximal entanglement production for spatial bipartitions,  see
\cite{dual-unitary-2,dual-gliders,max-velocity,tianci-max-velocity,bruno-generic-entanglement}.  Special cases of 
dual unitary operators are the so-called perfect tensors (with four legs), which lead to quantum  quantum circuits with maximal
mixing (see \cite{chaos-qchann,dual-unitary--bernoulli}).

Dual unitary matrices appeared much earlier in pure mathematics, namely in the study of planar algebras 
\cite{jones-planar}. In this context they were called bi-unitary matrices, see for example
\cite{biunitary-permutations}. Implications for quantum information theory were studied in \cite{biunitary-qinf1,biunitary-qinf2}.

Generalizations of dual unitary gates to other geometries also appeared in the literature. A generalization to the
triangular lattice was proposed in \cite{triunitary}.
The extension to  higher dimensional euclidean lattices with standard (hyper)-cubic geometry
was mentioned already in the seminal work \cite{dual-unitary-3}, and the case of the cubic lattice was worked out
recently in \cite{ternary-unitary}. The application of dual unitary gates in geometries with randomly scattered straight lines
was considered in \cite{prosen-mikado}.

{The concept of  ``absolutely maximally entangled state'' (AME) in quantum information theory is closely related to the concept of dual unitarity and its generalizations.} An AME is a multi-party state which has maximal bipartite entanglement for all
possible bipartitions of the system \cite{AME-1,AME-Helwig2,AME-Helwig3,AMEcomb1,four-AME}.
An AME is sometimes also called a ``perfect tensor''.
AMEs can be considered as a
quantum mechanical extension of orthogonal arrays known from combinatorial design theory \cite{OA-book}. Any
orthogonal array with the appropriate parameters can be used to construct a corresponding AME  \cite{AMEcomb1,AMEcomb2,AMEcomb3}, but the converse  
is not true. There are situations when the classical object (an orthogonal array of a given type) does not exist,
whereas the AME of the same type exists. A famous example is the problem of Euler's 36 officers
\cite{euler36,euler36-explanation}; for other similar cases see \cite{arul-perfect}. An online Table of AME can be found
at \cite{AME-list} (see also \cite{AME-bounds}).

AME can be used as quantum error correcting codes, and they appeared as 
 ingredients in the tensor network models of the AdS/CFT correspondence: they were used as holographic error
 correcting codes, see \cite{ads-code-1} and the recent review \cite{holocode-review}. 

{Using an operator-state correspondence, an AME state of four parties (i.e., a perfect tensor with four legs) can be interpreted as a special example of a dual unitary operator}. Similarly,
 AME with six or eight parties are  special examples of the generalizations of dual unitary operators considered in \cite{triunitary} and
\cite{ternary-unitary}. However, the requirements of an AME are much stronger than what is needed for many
applications. This led to the introduction of objects with weaker constraints: states that have maximal entanglement for
a limited set of bipartitions, such that the bipartitions are selected based on some geometrical principles.
In quantum information theory such objects appeared independently in
\cite{perfect-tangles,planar-AME}, and they were also introduced in the context of holographic error correcting codes in
\cite{block-perfect-tensor}. These generalizations involve planar arrangements of the parties of the state (legs of the
tensor), just like in the case of the tri-unitary gates of \cite{triunitary}. On the other hand, the gates of the work
\cite{ternary-unitary} correspond to a three dimensional arrangement of the constituents.

In order to capture all of the above generalizations of dual unitarity and maximal entanglement, we introduce the concepts of ``multi-directional unitarity'' and ``multi-directional maximal entanglement''. Our aim is to
give a general framework which encompasses all the objects mentioned above.

Among multi-directional unitary operators, our study focuses on those which are also completely invariant with respect to
the symmetry group of the given geometric arrangement. In the case of dual unitarity these are the operators which are
self-dual \cite{dual-unitary-3}; for other geometries the concept has not yet been investigated. In quantum information
theory similar works appeared recently, which studied quantum states with permutation invariance motivated by
geometrical symmetries, see for example \cite{Dicke-states-entanglement,Dicke-states-ent-2}. However, these works did
not investigate the states with maximal entanglement for the selected bipartitions.

The very recent work
\cite{huse-dual-clifford} considers so-called Clifford cellular automata, which are also multi-directional unitary. A
common point with our work is that \cite{huse-dual-clifford} also focuses on those cases where the arrangement possesses
exact geometrical symmetries. However, there is no overlap with our work, because they only consider qubit systems with
dual unitary Clifford gates, and they focus on the physical properties of the circuits. In contrast, we focus on the
constructions of the gates, or equivalently, the corresponding multi-leg tensors.

In Section \ref{sec:defs} we give all the detailed definitions, and specify the goals of this work. Afterwards,
Sections \ref{sec:du}-\ref{sec:classical} provide a number of constructions for the objects of our study.
We summarize our results in Section~\ref{sec:concl}, where we also present a short list of open problems.
Some additional computations are included in the Appendix.

\section{Multi-directional unitarity}

\label{sec:defs}

In this work we consider quantum states of product Hilbert spaces, and unitary operators acting on them. In all of our
cases we are dealing with homogeneous tensor products, which means that the actual Hilbert space is a tensor product of
finitely many copies of $\complex^N$ with some $N\ge 2$. 
We will work in a concrete basis.
Basis elements
of $\complex^N$ are denoted as $\ket{a}$ with $a=1,\dots,N$, and the basis elements of $\complex^N \otimes \complex^N$
are denoted simply as 
$\ket{ab}\equiv \ket{a}\otimes\ket{b}$. Extension to tensor products with more factors follows in a straightforward way.

First we discuss the dual unitary operators, which are the simplest and most studied multi-directional unitary
operators. Extensions to other geometries are given afterwards.

\subsection{Dual unitarity}

In order to introduce the dual unitary operators, we first consider unitary operators $\check U$ acting on the double
product $\complex^N\otimes\complex^N$. The notation $\check U$ is borrowed from the theory of
integrable models; we use $\check U$ here so that later we can also introduce another operator $U$ with a slightly
different geometrical interpretation.

The matrix elements of $\check U$ (or other two-site operators) will be denoted as $\check U_{ab}^{cd}$. We use the
convention
\begin{equation}
 \check  U\ket{ab}=\sum_{c,d=1}^N \check U_{ab}^{cd} \ket{cd}\,.
\end{equation}
The operator $\check U$ is unitary if
\begin{equation}
  \label{u1}
  \check U\check U^\dagger= \check U^\dagger\check U=1\,,
\end{equation}
which means in the concrete basis that
\begin{equation}
  \label{unitarity}
  \sum_{a,b=1}^N   \check U_{ab}^{cd}  (\check U_{ab}^{ef})^*=
\sum_{a,b=1}^N   \check U_{cd}^{ab}  (\check U_{ef}^{ab})^*
  =\delta^{ce}\delta^{df}\,.
\end{equation}
Here the asterisk means elementwise complex conjugation. The two equalities in \eqref{u1} or \eqref{unitarity}
are not independent, we write both of them
for the sake of completeness. 

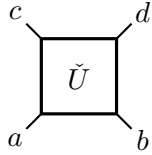
\begin{figure}[t]
  \centering
  \begin{tikzpicture}
    \draw [very thick] (0,0) rectangle ++(1,1);
    \draw [thick]  (0,0) to (-0.2,-0.2);
    \draw [thick]  (1,0) to (1.2,-0.2);
    \draw [thick]  (0,1) to (-0.2,1.2);
        \draw [thick]  (1,1) to (1.2,1.2);
        \node at (-0.35,-0.35) {$a$};
        \node at (1.35,-0.35) {$b$};
        \node at (-0.35,1.35) {$c$};
        \node at (1.35,1.35) {$d$};
\node at (0.5,0.5) {$\check U$};
      \end{tikzpicture} 
\caption{Pictorial representation of a dual unitary gate. In the standard interpretation $a$ and $b$ are the indices for
  the ``incoming'', and $c$ and $d$ for the ``outgoing'' spaces. Therefore, $\check U$ is seen as acting in the vertical
  direction upwards. In contrast, the reshuffled matrix $\check U^R$ acts from the left to the right, therefore its
  ``incoming'' indices are $c$ and $a$, and the outgoing indices are $d$ and $b$.
      } 
  \label{fig:du}
\end{figure}

Fig.~\ref{fig:du} shows a pictorial representation of a dual unitary gate. Here it is understood that the ``incoming'' and
``outgoing'' spaces (with indices $a,b$ and $c,d$) are
drawn as the lower and upper two legs of the gate, respectively. In this sense the picture represents time evolution in
the vertical direction, upwards.

The operator $\check U$ is dual unitary if it describes unitary evolution also in the space direction. This means that the reshuffled matrix $U^{R}$ given by the elements
\begin{equation}
  (\check U^R)_{ca}^{db}=\check U_{ab}^{cd}
  \label{eq:reshuffle}
\end{equation}
is also unitary. This second unitarity condition can be formulated directly for the matrix elements of the original operator
$\check U$, giving the relations (once again dependent on each other)
\begin{equation}
  \label{DU}
  \sum_{a,c=1}^N   \check U_{ab}^{cd}  (\check U_{ae}^{cf})^*=
 \sum_{a,c=1}^N   \check U_{ba}^{dc}  (\check U_{ea}^{fc})^*=
  \delta^{be}\delta^{df}\,.
\end{equation}

The constraints \eqref{unitarity}-\eqref{DU} define an algebraic variety. An explicit parametrization of this algebraic
variety is not known for $N\ge 3$. On the other hand, a complete description is available for $N=2$
\cite{dual-unitary-3}, see Section \ref{sec:du}. For higher dimensions only isolated constructions are known; 
an extensive review of known solutions was given in \cite{sajat-dual-unitary}. A numerical method to find new solutions was
presented in \cite{dual-ensembles-1}, for an early algebraic investigation see \cite{bipartite-unitaries}.

\subsection{Dual unitarity and highly entangled states}

\label{sec:duent}

In order to make the connection between dual unitary operators and highly entangled states, it is useful to treat the
dual unitary matrices as vectors in an enlarged Hilbert space. To achieve this, we define an operator-state
correspondence working in our concrete basis. 

Let us consider a four-fold tensor product space
\begin{equation}
  \HH=V_1\otimes V_2\otimes V_3\otimes V_4 \,,
\end{equation}
where $V_j\approx \complex^N$ for $j=1,\dots,4$.
We attach a geometrical interpretation to this space: we take a square and associate each space $V_j$ with a vertex of
the square. The ordering is such that the numbers $1, 2, 3, 4$ are put on the square in an anti-clockwise manner, see
Figure \ref{fig:squarehexa}.

Now we map every unitary operator $\check U$ to a state $\ket{\psi}\in \HH$. In our concrete basis
we choose the correspondence
\begin{equation}
  \label{opstate1}
  \psi_{abcd}=\frac{1}{N} \check U_{ab}^{dc}\,.
\end{equation}
The normalization is chosen such that $\skalarszorzat{\psi}{\psi}=1$.

It is well known that the different unitarity conditions imply that $\ket{\psi}$ has maximal entanglement for certain
bipartitions. For completeness we discuss this connection in detail.

We denote the set of sites of the chain as $\Sc=\{1,2,3,4\}$. 
Let us now consider the bipartition $\Sc=A\cup B$ with $A=\{1,2\}$ and $B=\{3,4\}$. The reduced density
matrix of the subsystem $A$ is given by
\begin{equation}
  \rho_A=\text{Tr}_B (\rho)=\text{Tr}_B \left( \ket{\psi}\bra{\psi}\right)=\text{Tr}_{3,4}   \left( \ket{\psi}\bra{\psi}\right)\,.
\end{equation}
By the definition of the partial trace the components are
\begin{equation}
  (\rho_A)_{ab}^{ef}=\sum_{c,d=1}^N \psi_{efcd}^{*} \psi_{abcd}.
\end{equation}
Combining this with \eqref{opstate1} and the unitarity condition \eqref{unitarity} we see that
\begin{equation}
   (\rho_A)_{ab}^{ef}=\frac{1}{N^2} \delta_{ae}\delta_{bf}.
\end{equation}
If the reduced density matrix is proportional to the identity then we speak of ``maximal entanglement'' or a ``maximally
mixed state''.  Indeed, in this
case all entanglement measures obtain their maximal values. For example, the von Neumann entropy becomes
\begin{equation}
  S_{vN}=-\text{Tr}_A \left[\rho_A\log(\rho_A)\right]=\log(N^2).
\end{equation}
Thus a unitary operator corresponds to a state having maximal entanglement between the pairs of
spins (or tensor legs) $\{1,2\}$ and $\{3,4\}$. 

The same computation can be performed also for the bipartition $\Sc=C\cup D$ with $C=\{1,4\}$ and $D=\{2,3\}$. In this
case the components of the reduced density matrix for subsystem $C$ become
\begin{equation}
  (\rho_C)_{ad}^{ef}=\sum_{b,c=1}^N \psi_{ebcf}^{*}\psi_{abcd}.
\end{equation}
Comparing again with \eqref{opstate1} we see that if the dual unitary condition \eqref{DU} holds, then
\begin{equation}
   (\rho_C)_{ad}^{ef}=\frac{1}{N^2} \delta_{ae}\delta_{df}.
\end{equation}
Therefore, we again obtain maximal entanglement for the given bipartition. 

\subsection{Absolutely maximally entangled states}

Absolutely maximal entanglement is a concept from quantum information theory, which is related to the property of dual
unitarity described above. In order to define it, let us consider a generic 
Hilbert-space of $K$ qudits:
\begin{equation}
  \label{HNK}
  \HH=\otimes_{j=1}^K V_j,\qquad V_j\approx \complex^N.
\end{equation}
We denote the corresponding set of sites as $\Sc=\{1,2,\dots,K\}$, and for simplicity we assume that $K$ is even. A
 state $\ket{\psi}\in \HH$ is absolutely maximally entangled (AME) if it has maximal entanglement for
every bipartition of the system. More precisely, $\ket{\psi}$ is an AME if for any bipartition $A \cup B = \Sc$
  with $|A| \le |B|$, the reduced density matrix $\rho_A=\text{Tr}_B (\ket{\psi}\bra{\psi})$ is
proportional to the identity matrix. The set of such states is generally denoted as $\text{AME}(K,N)$. 
AMEs have been discussed in the literature in various works; selected references are
\cite{AME-1,AME-Helwig2,AME-Helwig3,AME-bounds,AMEcomb1,AMEcomb2,AMEcomb3},
for an online
table see \cite{AME-list}. An AME is also called a ``perfect tensor''  \cite{chaos-qchann}. The unitary operators that
are obtained from AME via the operator-state correspondence were called multi-unitary \cite{AMEcomb2}, and specializing
to cases with $K=4,6,\dots$ they were called 2-unitary, 3-unitary, etc.

In the computations of the previous subsection we saw that a state $\ket{\psi}$ corresponding to a DU operator satisfies
some part of the 
conditions of being an AME. In that case we have $K=4$ and the maximal entanglement criterion holds for the bipartitions
$\{1,2\}\cup \{3,4\}$ and $\{1,4\}\cup\{2,3\}$. However, there is no condition for the bipartition $\{1,3\}\cup\{2,4\}$,
which corresponds to ``isolating'' to the two diagonals of the square (see Figs. \ref{fig:du} and \ref{fig:squarehexa}).
Unitary operators $U$ for which maximal entanglement also holds for the diagonal bipartition were called
``Bernoulli'' in \cite{dual-unitary--bernoulli}.

\subsection{Multi-directional unitarity}

Generalizations of the “square-shaped” (see Fig. \ref{fig:du}) dual unitary operators to more complicated geometries such as hexagons or cubes have appeared in various places in the literature  but an overall framework to describe all the various cases is lacking. In the following, we outline a general framework for these generalizations which we call ``multi-directional unitary operators''.

Once again we take $K$ copies of the Hilbert space $\complex^N$, so
the full Hilbert space is given by \eqref{HNK}.
In all our cases $K$ is an even number.
We consider vectors $\ket{\psi}\in \HH$ with special entanglement properties. The idea is that the states should have
maximal entanglement for selected bipartitions, and the selection is motivated by certain geometric arrangements of
the ``parties'' or sites.

The condition of maximal entanglement is the same as in the dual unitary case. Let us take a bipartition $S=A\cup B$ such that $|A|=|B|=K/2$. Once again we say that there is maximal entanglement
between 
subsystems $A$ and $B$ if
\begin{equation}
  \rho_A=\text{Tr}_B(\rho) \sim 1_A\,,\qquad \text{ and }\qquad  \rho_B=\text{Tr}_A(\rho) \sim 1_B.
  \label{eq:multimax}
\end{equation}
The two conditions imply each other. If \eqref{eq:multimax} holds, then the vector $\ket{\Psi}$ can be interpreted as a unitary
operator $\HH_A\to \HH_B$, where $\HH_A$ and $\HH_B$ are the two Hilbert spaces corresponding to the two subsystems $A$ and
$B$. The precise form of this operator-state correspondence is given below in Section \ref{sec:opstate}.

The generalization of dual unitarity from the square to other geometries is the following: We arrange the sites $1,\dots,K$ into a
symmetric geometric pattern in some Euclidean space. In most of our examples  we are dealing
with planar arrangements but we also treat some three-dimensional arrangements. 
Based on the chosen arrangement, we select a list of non-empty
  subsets $\{A_1,A_2,A_3,\dots,A_{n}\}$, $A_j\subset \Sc$, and require that the state $\ket{\psi}$ is maximally
entangled for all bipartitions
\begin{equation}
  \label{bipartitions}
  \Sc=A_j\cup (\Sc\setminus A_j),\qquad j=1,\dots, n
\end{equation}
The selection of the subsets follows from the geometric symmetries of the spatial arrangement. In all cases
the bipartition is obtained by cutting the set $\Sc$ into two equal parts in a symmetric way.

We suggest to call the states that satisfy the above requirements 
``multi-directional maximally entangled states'', and to call the corresponding operators ``multi-directional unitary operators''.
For the operators, the name ``multi-unitary operator'' would also be appropriate, but that has been already used in the context of the AME states
\cite{AMEcomb2,AME-review}.

Below we give a list of the geometric arrangements that we consider in this article. Most of these arrangements have already appeared in the literature, which provides motivation to study them.
\begin{itemize}
\item {\bf Square.} This corresponds to the dual unitary operators discussed above. In this case $K=4$, the four sites are
  arranged as the four vertices of a square (see Fig. \ref{fig:squarehexa}), and the subsets $A_j$ are 
  the pairs of vertices on the four edges of the square. Therefore the allowed bipartitions correspond to cutting the
  square into two halves such that both halves have a pair of neighbouring vertices.

Dual unitary gates can be used to build quantum cellular automata in 1+1 dimensions. They are the simplest and most
studied examples of the multi-directional unitary operators.
\item {\bf Hexagon.} In this case $K=6$ and the sites are arranged as the vertices of a regular hexagon (see Fig. \ref{fig:squarehexa}). The allowed
  bipartitions correspond to cutting the hexagon into two equal parts, such that both parts contain three neighbouring
  sites.
The case of the hexagon was
   treated in \cite{triunitary} and the resulting operators were called ``tri-unitary operators''.

   The resulting operators can be used to build quantum cellular automata in 1+1 dimensions using the geometry of the
   triangular lattice \cite{triunitary}. 
\item {\bf Regular polygons.} In this case the sites are arranged in a plane as the vertices of a regular polygon with
  $K=2k$ sides; this generalizes the previous cases of the square and the hexagon. The allowed subsets $A_j$ consist of
  the consecutive neighbouring $k$-tuplets (with periodic boundary conditions over the set of sites $1,\dots,K$). The
  bipartitions are performed along symmetry axes that do not include 
  any of the vertices. States that satisfy maximal entanglement in these arrangements appeared independently in (at
  least) three works in the literature:
  as ``perfect tangles'' in \cite{perfect-tangles},  as ``block perfect tensors'' in \cite{block-perfect-tensor}
and as ``planar maximally entangled states'' in \cite{planar-AME}. See also   \cite{planar-OA} where the corresponding
classical object was called ``planar orthogonal array''.
\item {\bf Cube.} This is our first example where the sites are arranged in a three dimensional setting. Now we have
  $K=8$ sites which are put
  onto the vertices of a cube (see Fig.~\ref{fig:cube}). The allowed subsets $A_j$ are the sets of four vertices belonging to one of the faces of
  the cube. Therefore, the bipartitions are obtained by cutting the cube into two halves along symmetry planes that
  are parallel to the faces.

  This case was considered very recently in \cite{ternary-unitary}, the resulting
  operators were called ``ternary unitary''. They can be used to build quantum cellular automata in 2+1 dimensions, in
  the arrangement of the standard cubic lattice.

\item {\bf Octahedron.} This is also an example in 3D. Now we have $K=6$ sites which are arranged as the six vertices of
  an octahedron, or alternatively, as the six faces of a cube (see Fig.~\ref{fig:cube}). The bipartitions are obtained by cutting the
  octahedron into two equal parts, each of them containing 3 vertices. Alternatively, if the vertices of the octahedron
  are identified as the faces of a 
  cube, the allowed triplets of sites are obtained by selecting three faces of the cube which join each other in a
  chosen vertex. Note that the hexagonal arrangement also has $K=6$, but the set of allowed bipartitions is 
  different in the two different geometries.  Below we show that the octahedral case is actually a specialization of the
  hexagonal case: if proper identifications are made, then the octahedral case is seen to have one more unitarity
  condition on top of those of the hexagonal case.

  The octahedral arrangement can also be used to build quantum cellular automata in 2+1 dimensions. The idea is to take the
  standard cubic lattice, put the local spaces on the faces of the cubic lattice, and to 
 interpret a body diagonal as the direction of time.  In this case the time slices are given
 by kagome lattices.
 The work  \cite{triunitary} used this geometry, but with gates that come from the hexagonal
 arrangement.
\item {\bf Tetrahedron.}  We add this case for completeness, even though it does not describe a new structure.
  Now we have
  $K=4$, and the four sites are put onto the four vertices of a tetrahedron. The allowed bipartitions are the three
  possible ways 
  to cut the four sites into two pairs. Therefore the resulting objects are the AMEs of four parties.
\end{itemize}

In the list above we did not include the remaining two Platonic solids, the dodecahedron and the icosahedron. This is because we do not know of any application of states/operators with these geometries. However, the extension of the
definition to these geometries is straightforward.

\subsection{Identity operators and operator-state correspondence}

\label{sec:opstate}

{The definition of the partial trace \eqref{eq:multimax} suggests the idea to use identity operators
  as multi-directional unitary operators. In such cases the reduced density matrices always become identity matrices
  themselves (although of smaller size). However, there are various options for the operator-state correspondence, and not
  all of them yield multi-directional unitary operators. For example, in the case of the square geometry, $\check U_{ab}^{cd}=\delta^{ac}\delta^{bd}$ and  $\check U_{ab}^{cd}=\delta^{ad}\delta^{bc}$ both seem to be legitimate identity operators. However, one can easily check that only the latter one is an actual dual unitary operator.
}

In the following, we use a symmetry consideration to obtain the identity operator that is multi-directional unitary. In
all of the geometric arrangements considered above 
(except the tetrahedron), each vertex has a well defined
``opposite vertex'', or antipode, to which it is connected by a diagonal. The total number of 
diagonals that connect antipodes is $K/2$. Each such
diagonal includes exactly two vertices of the arrangement, and for each allowed
bipartition $A_j\cup (\Sc\setminus A_j)$ exactly one vertex is included in $A_j$ from each such diagonal. In the case of the
regular polygons these diagonals are the 
standard diagonals that connect opposite points,
whereas in the case of the cube and the octahedron they are the
space diagonals. We choose the identity operator to be that which acts as identity along these diagonals.
This operator is invariant under the symmetry group of the geometrical arrangement, and therefore a multi-directional
unitary operator. 

In the following we explicitly write down the identity operators and the corresponding states in the various geometries. By doing this, we also set the convention
of the operator-state correspondence in each case.
Based on this operator-state correspondence, an identity operator described above corresponds to a state made up as the product of $K/2$ maximally entangled Bell pairs prepared on the pairs of opposite vertices.

\begin{figure}[t]
  \centering
  \begin{tikzpicture}
    \draw [thick] (0,0) -- (0,2) -- (2,2) -- (2,0) -- (0,0);
    \node at (-0.2,-0.2) {$1$};
    \node at (2.2,-0.2) {$2$};
    \node at (2.2,2.2) {$3$};
    \node at (-0.2,2.2) {$4$};

    \begin{scope}[xshift=6cm]
   \draw [thick] (0,0) -- (1,0) -- (1.5,0.8660) -- (1,1.7320) -- (0,1.7320) -- (-0.5,0.8660) -- (0,0);
    \node at (-0.1,-0.2) {$1$};
    \node at (1.1,-0.2) {$2$};
    \node at (1.7,0.8660) {$3$};
    \node at (-0.1,1.9320) {$5$};
    \node at (1.1,1.9320) {$4$};
    \node at (-0.7,0.8660) {$6$};      
    \end{scope}
    
  \end{tikzpicture}
  \caption{Labels for the local vector spaces in the square and the hexagonal geometries. The simplest multi-directional unitary
    operators are those  
    which act identically along the diagonals. In the case of the square they correspond to maximally entangled two-site
    states prepared on the pairs of sites $(1,3)$ and $(2,4)$. For the hexagon they correspond to the Bell pairs on the pairs
of sites $(1,4)$, $(2,5)$ and $(3,6)$.}
  \label{fig:squarehexa}
\end{figure}
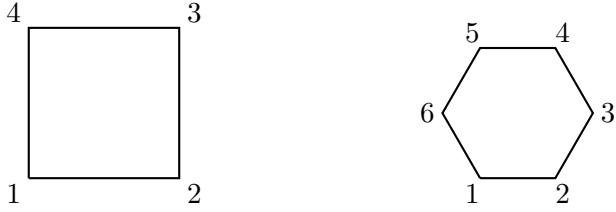

\bigskip

{\bf Square geometry.} 
This case was already treated above, but it is useful to repeat the operator-state
correspondence. Now there are $K=4$ sites. The local Hilbert
spaces will be indexed by the labels
$1, 2, 3, 4$, which are associated with the four vertices of the
square; we write them down in an anti-clockwise manner (see Fig.
\ref{fig:squarehexa}). Now the operator-state correspondence for $\check U$ is introduced via formula
\eqref{opstate1}. It can be seen that  $\check U$  acts along two edges of the square. On the other hand, we also
introduce the operator $U$ via
\begin{equation}
  \label{opstate2}
  \psi_{abcd}=\frac{1}{N} U_{ab}^{cd}.
\end{equation}
It can be seen that $U$ acts along the diagonals of the square.  The two operators are connected by the relation $\check
U=\PP U$, where $\PP$ is the two-site permutation operator (SWAP gate).

So far the dual unitarity conditions were formulated for the
elements of the matrix
$\check U$ \eqref{DU}, now we also consider the conditions for the
elements of
$U$. It follows from the above correspondences, that $U$ is dual unitary if both  $U$ and $ U^{t_1}$ are 
unitary, where $t_1$ means partial transpose with respect to the first space. As an equivalent requirement, $U$ and
$U^{t_2}$ should 
be both unitary.

As explained above, the operator $U=1$ is dual unitary, and it corresponds to the state
\begin{equation}
  \ket{\psi}=\frac{1}{N}\sum_{a,b=1}^N \ket{abab}.
\end{equation}
This is a tensor product of two Bell pairs prepared on the two
diagonals of the square.

\bigskip

{\bf Hexagonal geometry.}
Now the $K=6$ local spaces are associated with the 6 vertices of a hexagon; we
put the labels anti-clockwise onto the vertices (see Fig.
  \ref{fig:squarehexa}). The standard operator-state
correspondence is through the 
relation
\begin{equation}
 \psi_{abcdef} = \frac{1}{\sqrt{N^3}}  \check U^{fed}_{abc}.
\end{equation}
This corresponds to the definitions in \cite{triunitary}.
Alternatively, we use the correspondence
\begin{equation}
  \psi_{abcdef}= \frac{1}{\sqrt{N^{3}}}     U_{abc}^{def}.
\end{equation}
The connection between the two definitions is given by
\begin{equation}
  U=\PP_{1,3}\check U,
\end{equation}
where $\PP_{1,3}$ permutes spaces 1 and 3.
In these conventions an operator $U$ is multi-directional unitary if
  $U$, $U^{t_1}$ and also $U^{t_3}$ are unitary. 

The state corresponding to the identity operator $U=1$ is given by
\begin{equation}
  \ket{\psi}=\frac{1}{N^{3/2}}\sum_{a,b,c=1}^N \ket{abcabc}.
\end{equation}
This is a product of three Bell pairs prepared on the long
  diagonals of the hexagon.

\begin{figure}[t]
  \centering
  \begin{tikzpicture}[every edge quotes/.append style={auto}]
  \pgfmathsetmacro{\cubex}{2}
  \pgfmathsetmacro{\cubey}{2}
  \pgfmathsetmacro{\cubez}{2}
  \draw [thick, every edge/.append style={densely dashed, opacity=.5}]
    (0,0,0) coordinate (o) -- ++(-\cubex,0,0) coordinate (a) -- ++(0,-\cubey,0) coordinate (b) edge coordinate [pos=1] (g) ++(0,0,-\cubez)  -- ++(\cubex,0,0) coordinate (c) -- cycle
    (o) -- ++(0,0,-\cubez) coordinate (d) -- ++(0,-\cubey,0) coordinate (e) edge (g) -- (c) -- cycle
    (o) -- (a) -- ++(0,0,-\cubez) coordinate (f) edge (g) -- (d) -- cycle;
    \node at (0.2,-0.2,0) {$8$};
    \node at (-2.2,-0.2,0) {$7$};
    \node at (-2.2,-2.2,0) {$1$};
    \node at (0.2,-2.2,0) {$2$};
      \node at (0.2,-0.2,-2) {$5$};
    \node at (-2.2,-1.8,-2) {$4$};
    \node at (-2.2,0.2,-2) {$6$};
    \node at (0.2,-2.2,-2) {$3$};

    \begin{scope}[xshift=5cm,yshift=-0.6cm]
      \coordinate (A1) at (0,0);
\coordinate (A2) at (1.8,0.6);
\coordinate (A3) at (3,0);
\coordinate (A4) at (1.2,-0.6);
\coordinate (B1) at (1.5,1.5);
\coordinate (B2) at (1.5,-1.5);

\node at (-0.2,0) {$1$};
\node at (1.95,0.75) {$6$};
\node  at (3.2,0)  {$4$};
\node  at (1.05,-0.75)  {$3$};
\node  at (1.5,1.7)  {$5$};
\node  at (1.5,-1.7)  {$2$};

\begin{scope}[thick,dashed,opacity=0.6]
\draw (A1) -- (A2) -- (A3);
\draw (B1) -- (A2) -- (B2);
\end{scope}
\draw[thick] (A1) -- (A4) -- (B1);
\draw[thick] (A1) -- (A4) -- (B2);
\draw[thick] (A3) -- (A4) -- (B1);
\draw[thick] (A3) -- (A4) -- (B2);
\draw [thick] (B1) -- (A1) -- (B2) -- (A3) --cycle;
    \end{scope}
    
  \end{tikzpicture}
  \caption{Labels for the vector spaces in the cubic and octahedral geometries. The simplest multi-directional unitary
    operators are those  which act as the 
  identity matrix along the body diagonals. In terms of the quantum states this corresponds to preparing maximally entangled
  two-site states on the pairs of sites $(1,5)$, $(2,6)$, $(3,7)$ and $(4,8)$ in the cubic, and $(1,4)$, $(2,5)$ and
  $(3,6)$ in the octahedral geometries.}
  \label{fig:cube}
\end{figure}
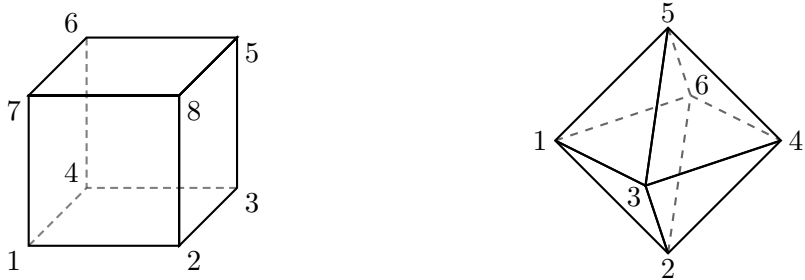

\bigskip
{\bf Cubic geometry.}
Now we have $K=8$ sites, and we introduce the following labelling for them: the sites $1, 2, 3, 4$
are placed on a selected face in the given order, and the sites $5, 6, 7, 8$ are on the other face, such that the sites
$j$ and $j+4$ with $j=1,\dots,4$ are on the same body diagonal
(see Figure \ref{fig:cube}). Again there are two
possibilities for an operator-state correspondence. We introduce the matrix $\check U$ which acts from the tensor
product of the spaces $1, 2, 3, 4$ to the product of spaces $5, 6, 7, 8$, such that the ``incoming'' and ``outgoing''
spaces are connected by the edges of the cube. This gives the operator-state correspondence
\begin{equation}
{\psi_{a_1a_2a_3a_4a_5a_6a_7a_8}=  \frac{1}{N^{2}} \check
U^{a_7a_8a_5a_6}_{a_1a_2a_3a_4},  }   \qquad a_j=1,\dots,N, \qquad
j=1,\dots,8.
\end{equation}
The $\check U$ operator is used to build the quantum circuits in 2+1 dimensional space, see \cite{ternary-unitary}. On
the other hand, we again introduce an operator $U$ which acts along
the body diagonals. In this case the operator-state
correspondence is
\begin{equation}
{\psi_{a_1a_2a_3a_4a_5a_6a_7a_8}=\frac{1}{N^{2}}
U^{a_5a_6a_7a_8}_{a_1a_2a_3a_4}},\qquad a_j=1,\dots,N, \qquad
j=1,\dots,8.
\end{equation}
In these conventions an operator $U$ is multi-directional unitary, if $U$, $U^{t_2t_3}$ and $U^{t_3t_4}$ are all unitary.

{The identity operator can be constructed in the same way as for the
square and the hexagonal geometries. The corresponding state will be a
product of four Bell pairs over the space diagonals of the cube.}

\bigskip

{\bf Octahedral geometry.} Now we have $K=6$ sites, and we label them according to the pattern shown in
Fig. \ref{fig:cube}. Now the pairs $(1,4)$, $(2,5)$ and $(3,6)$ are the body diagonals. In accordance, we introduce the
operator-state correspondence as
\begin{equation}
  \psi_{abcdef}=\frac{1}{{\sqrt{N^{3}}}} U_{abc}^{def}.
\end{equation}
 In these conventions an operator $U$ is multi-directional unitary, if $U$, $U^{t_1}$, $U^{t_2}$ and $U^{t_3}$ are
all unitary.

Let us compare these conditions with those of the hexagonal case
where we also have $K=6$.
We see that the octahedral conditions are stronger: they include one extra condition on top of those of the hexagonal
case. This extra condition is the unitarity of $U^{t_2}$. It follows that every octahedral unitary matrix is hexagonal
unitary, but the converse is not true. The similarity between these constraints was used in \cite{triunitary}, where a
2+1 dimensional quantum circuit was built using the hexagonal unitary gates. 

In the octahedral case, the identity operator can be constructed in the same way as for the
square and the hexagonal geometries. The corresponding state will be a
product of three Bell pairs over the diagonals of the octahedron.

\subsection{Symmetries and equivalence classes}

Let us discuss the operations on the states $\ket{\psi}$ which preserve the required entanglement properties. These operations will be used to obtain classifications of the various multi-directional unitary operators.

Among the operations, we distinguish two classes.
The first class is the class of local unitary operations under which all entanglement measures are
invariant. Consequently, if $\ket{\psi}$ 
satisfies the constraints of multi-directional unitarity in one of the geometries, then the state
\begin{equation}
  \label{LU}
  (U^{(1)}\otimes U^{(2)}\otimes \dots\otimes U^{(K)})\ket{\psi},\qquad U^{(j)}\in SU(N),\quad
  j=1,2,\dots,K
\end{equation}
also satisfies them. If two vectors $\ket{\psi_1}$ and $\ket{\psi_2}$ can be transformed into each other in this way
then we say that they are local unitary (LU) equivalent.

The second class of operations consists of those
spatial rearrangements (permutations)
of the sites which respect the geometry at hand. We denote the
corresponding symmetry group as $G$. In
the case of regular polygons the spatial symmetries are given by the dihedral group $G=D_K$, whereas for the cube and the
octahedron the spatial symmetries are given by the octahedral group
$G=O_h\sim S_4\otimes Z_2$. Every symmetry transformation $g\in G$ can be understood as a permutation of the sites,
which leads to a permutation operation $\PP_g$ on the tensor product of the individual vector spaces.  It is clear that
if  $\ket{\psi}$ satisfies the entanglement criteria, then $\PP_g \ket{\psi}$ also satisfies them for every $g\in G$.

We say that two states $\ket{\psi_1}$ and $\ket{\psi_2}$ are 
equivalent if they can be transformed into each other using
a combination of a permutation $\PP_g$ and local unitary operations.

\subsection{Spatially symmetric states}

\label{sec:spatially}

In this work we consider those multi-directional maximally entangled states (or multi-directional unitary operators)
that are also maximally symmetric with respect to the 
corresponding spatial symmetry
group. This means that for any $g\in G$ the states satisfy
\begin{equation}
  \label{psym}
  \PP_g\ket{\psi}=\ket{\psi}.
\end{equation}
In this work we call these states ``spatially symmetric''.
The spatially symmetric states have the appealing property that their concrete representation does not depend on a
chosen orientation of 
the geometric arrangement.
Furthermore, if we require the symmetry \eqref{psym} then the property of maximal entanglement
has to be checked only 
in one of the allowed bipartitions.
In practice this means that one needs to check 
the unitarity of only one matrix, because the matrix of the operator remains the same under all the spatial symmetry
transformations.

Once the permutation symmetry \eqref{psym} is required, the general local unitary transformations \eqref{LU} are not
allowed anymore. However, the symmetric unitary transformations
\begin{equation}
  \label{LU2}
\ket{\psi}\quad\to\quad  (V\otimes V\otimes \dots\otimes V)\ket{\psi},\qquad V\in SU(N)
\end{equation}
are still allowed. We say that two spatially symmetric states $\ket{\psi_1}$ and $\ket{\psi_2}$ are equivalent if there is a
one-site operator $V$ which transforms them into each other. 

Our requirement \eqref{psym} for the spatial symmetry is rather strong, because it requires invariance in a strict
sense. We could also require the weaker condition that for each $g\in G$ the state $\PP_g\ket{\psi}$ should be LU
equivalent to $\ket{\psi}$. For completeness we show two examples for this in Appendix \ref{app:luperm}.

In the case of dual unitary matrices certain solutions with other types of symmetry properties have already been
considered in the literature. Perfect tensors with $SU(2)$ invariance were studied in \cite{invariant-perfect}, whereas
dual unitary matrices with diagonal $SU(N)$ or diagonal $SO(N)$ symmetries were analyzed in
\cite{diagonal-du-1,diagonal-du-2,diagonal-du-3}. However, the spatial symmetry that we consider has not yet been
investigated in the context of these maximally entangled states.

\subsection{Further entanglement properties}

The definition of multi-directional unitarity enforces maximal entanglement along selected bipartitions. States
satisfying these requirements can still have very different bipartite or multi-partite entanglement patterns. 

As it was explained above, the identity operations along the diagonals are multi-directional unitary. They correspond
to states with maximally entangled pairs prepared on the diagonals. In these states the diagonals are completely
independent from each other: there is no entanglement between them. To be more precise: if we group the two sites on each
diagonal into a macro-site of dimension $N^2$, then we obtain a product state.
In the quantum circuits that are built out of these objects \cite{dual-unitary-3,triunitary,ternary-unitary} the
identity matrices describe free information propagation in selected directions of the circuit. Interaction happens in
the circuit only if the diagonals of the state are entangled with each other. 

Therefore it is very natural to consider the bipartite or multi-partite entanglement entropies {\it between the
  diagonals}. Non-zero entanglement between the diagonals signals that the state $\ket{\psi}$ is non-trivial, and the
corresponding operator $\check U$ generates interactions in the quantum circuits.  In this work we focus on these
non-trivial cases. 

\section{Dual unitary matrices}

\label{sec:du}

In the previous Section, we have laid the general framework for treating multi-directional unitary operators. We have also restricted our attention to operators that are invariant under the spatial symmetries of the arrangement. In the rest of the article, we discuss their various special cases.

First we consider the square geometry and dual unitary matrices.
Invariance with
respect to the geometrical symmetry operations means that 
the matrix of $U$ coincides with the 7 reshuffled matrices that arise from the
7 non-identical transformations of the square. The symmetry group is generated by a rotation and a reflection, therefore it
is enough to check invariance with respect to those symmetries.

For the  matrix $\check U$ these two transformations lead to
\begin{equation}
  \check U\to \check U^R,\qquad \check U\to \PP U \PP,
\end{equation}
where the definition of $\check U^R$ is given by \eqref{eq:reshuffle}. Therefore, a symmetric matrix is one which satisfies
\begin{equation}
\label{dusym1}
  \check U=\check U^R=\PP U \PP.
\end{equation}
In terms of the matrix $U$ the two necessary relations can be chosen as
\begin{equation}
  \label{dugeom1}
   U = U^{t_1}=\PP  U \PP
\end{equation}
Matrices that satisfy the properties 
\eqref{dusym1} or \eqref{dugeom1} are called self-dual.

Note that our conventions imply that the time reflection
step does not involve any complex conjugation, and it is given merely by the full transpose of the operators. This is to
be contrasted with the usual choice in quantum mechanics, where time reflection also involves a complex conjugation.
However, our choice is natural here, because we are treating a space-time duality, and time reflection in a
given direction describes space reflection in the other direction, therefore it is most natural that it does not involve
a complex conjugation.

The symmetric unitary transformations \eqref{LU2} result in
\begin{equation}
  U\quad\to \quad (V^t\otimes V^t) U (V\otimes V).
  \label{eq:vvuvv}
\end{equation}
The transposition in the factors on the left follows simply from the operator-state correspondence, and 
the symmetric prescription \eqref{LU2} for the transformation of the states.

\subsection{Two qubits}

Let us consider the special case of $N=2$, i.e., operators acting on two qubits. In this case, dual
 unitary gates were completely classified  in \cite{dual-unitary-3} based on the Cartan decomposition of a general
$U(4)$ matrix. It was found in \cite{dual-unitary-3}  that a generic dual unitary matrix can be written as
\begin{equation}
  \label{cartan1}
   U=e^{i\varphi}(S_1\otimes S_2) D(S_3\otimes S_4),
\end{equation}
where $S_{1,2,3,4}\in SU(2)$ are one-site operators, $\varphi\in\valos$, and $D$ is a diagonal unitary matrix, for which
a parametrization
can be chosen as
\begin{equation}
  D=e^{i \alpha Z_1 Z_2}=
\cos(\alpha)+i\sin(\alpha) Z_1Z_2=
  \text{diag}(e^{i\alpha},e^{-i\alpha},e^{-i\alpha},e^{i\alpha}),\qquad \alpha\in\valos.
\end{equation}
Note that we gave the representation of the matrix $U$, and not $\check U$.

Now we consider the self-dual cases. First of all we observe that the matrix $D$ is self-dual for every $\alpha$: it is
easy to see that it satisfies the two conditions of \eqref{dugeom1}. From
the form \eqref{eq:vvuvv} of the self-duality preserving transformations
it follows that self-dual solutions are
produced as
\begin{equation}
  \label{du2a}
 e^{i\varphi} (V^t\otimes V^t) D(V\otimes V), \qquad V\in SU(2).
\end{equation}
This gives a large family of self-dual solutions, with a total number of 5 real parameters: 3 parameters of $V$, the
``inner phase'' $\alpha$ and the ``global phase'' $\varphi$.

It can be shown that \eqref{du2a} covers all
self-dual
cases. Every dual unitary matrix with $N=2$ is of the
form \eqref{cartan1}, and the decomposition is unique if $\alpha\ne 0$.
The spatial symmetry operations
simply just exchange the external one-site operators, therefore the invariance property implies the form
\eqref{du2a}. In the special case $\alpha=0$ we are dealing with a simpler situation, namely $U=(S_1S_3)\otimes
(S_2S_4)$.
Such a decomposition is clearly not unique, but in
the case of $U$ being self-dual, the form \eqref{du2a} can be established nevertheless.

A very important special case is the evolution matrix of the {\it kicked Ising model} at the self-dual point
\cite{kicked-ising-eredeti,kicked-Ising-space-time-duality-1,sarang-lamacraft-kicked-DU,dual-unitary-1,dual-unitary-2,dual-kicked}. It
can be written as
\begin{equation}
  \check U=e^{i\frac{\pi}{4}Z_1Z_2}e^{i\frac{\pi}{4}(X_1+X_2)}e^{i\frac{\pi}{4}Z_1Z_2}.
\end{equation}
Note that now we gave the operator $\check U$.
In a concrete matrix representation we have
\begin{equation}
  \label{kicked2}
\check U=\frac{i}{2}\begin{pmatrix}
  1  & -1 & -1 & -1 \\
  -1 & -1 & 1 & -1 \\
  -1 & 1 & -1 & -1\\
  -1 & -1 & -1 &1
\end{pmatrix} .
\end{equation}
This is proportional to a real Hadamard matrix.
We will return to this specific matrix later, because it is the simplest example
for two general constructions that we treat in Sections \ref{sec:had} and \ref{sec:graph}.

\section{Diagonal unitary matrices}

The simplest idea for multi-directional unitary operators is to take the identity matrices (with the
proper operator-state correspondence laid out above) and to dress them with phase factors. For the dual unitary matrices
this idea was 
explored in detail in \cite{dual-unit-param}, whereas in the hexagonal geometry
it appeared in \cite{triunitary}. Here we also discuss this construction, by focusing on the spatially symmetric
solutions. Below we give the details in the square and hexagonal geometries; the extensions to the cube and the
octahedron are straightforward.

\subsection{Dual unitary matrices}

In the case of dual unitary matrices (square geometry) we take $U=D$ with $D$ being a diagonal unitary matrix with
the non-zero elements being
\begin{equation}
  \label{Dab}
 D_{ab}^{ab}= e^{i\varphi_{ab}},\qquad \varphi_{ab}\in\valos, \quad a,b=1,\dots,N.
\end{equation}
As first explained in \cite{dual-unit-param}, such a matrix is dual unitary for arbitrary phases and arbitrary $N$. This
follows immediately from the fact that the partial transpose of a diagonal matrix is itself. 

An alternative ``physical interpretation'' can be given as follows: the gate $D$ describes a scattering event of two particles on two
crossing world lines. The two particles have an inner degree of freedom, which is indexed by the labels
$a,b=1,\dots,N$. The incoming and outgoing particles have the same label, 
and there is a phase factor associated with each event. This phase depends on the pairs of labels $(a,b)$. 

In order to obtain a spatially symmetric (self-dual) solution one needs to check the symmetry condition, which is the
second relation in \eqref{dugeom1}. In this concrete case one obtains the simple symmetry relation
\begin{equation}
  \varphi_{ab}=\varphi_{ba}.
\end{equation}
Thus we obtain a large family of self-dual solutions, with the number of real parameters being $N(N+1)/2$. However, we
can set the phases $\varphi_{aa}$ to zero with symmetrically placed one-site operators, therefore the total number of
intrinsically independent phases is $N(N-1)/2$.  

For completeness we also give the four site state  corresponding to \eqref{Dab}. It reads
\begin{equation}
  \ket{\psi}=\frac{1}{N}\sum_{a,b=1}^N e^{i\varphi_{ab}}\ket{a,b,a,b}.
\end{equation}
If $\varphi_{aa}=0$ but $\varphi_{ab}\ne 0$ for
some $a,b$, then these states have entanglement of the diagonals,
therefore they generate interactions in the quantum circuits. The effects of these interactions were explored in detail
in \cite{dual-unit-param}.

\subsection{Hexagonal geometry}

We use the notations introduced in Section \ref{sec:opstate} for the case of the hexagonal geometry. In this case we
treat a diagonal matrix $U=D$, where now the matrix elements are
\begin{equation}
  \label{Dabc}
  D_{abc}^{abc}= e^{i\varphi_{abc}},\qquad \varphi_{abc}\in\valos, \quad a,b,c=1,\dots,N.
\end{equation}
It is easy to see that a full hexagonal symmetry is achieved only if $\varphi_{abc}$ is completely
symmetric with respect to $a,b,c$. Generally this gives a large family of solutions. For completeness we also give the six site state  corresponding to \eqref{Dabc}. It reads
\begin{equation}
  \ket{\psi}=\frac{1}{N^{3/2}}\sum_{a,b,c=1}^N e^{i\varphi_{abc}}\ket{abcabc}.
\end{equation}
In the case of qubits ($N=2$) the solutions can be parametrized as 
\begin{equation}
  \label{D3Z}
  D=\exp\left[i(\alpha+\beta(Z_1+Z_2+Z_3)+\gamma(Z_1Z_2+Z_2Z_3+Z_1Z_3)+\delta Z_1Z_2Z_3)\right].
\end{equation}
The parameters $\alpha$ and $\beta$ control a global phase and three one-site unitary operations, and we are free to set
them to 0. The remaining ``inner'' parameters are $\gamma$ and $\delta$.
If $\delta=0$ then the three site gate factorizes as the product of two-site operators 
\begin{equation}
  D_{123}=E_{12}E_{13}E_{23},\qquad  E_{jk}=e^{i\gamma Z_jZ_k}.
\end{equation}
However, if $\delta\ne 0$ then a true three-body interaction term appears, which couples the three diagonals of the
hexagon. For $\gamma\ne 0$ and/or $\delta\ne 0$ there is non-zero entanglement between the diagonals of the hexagon.

\section{Constructions from Hadamard matrices}

\label{sec:had}

It is known that dual unitary matrices can be constructed from complex Hadamard matrices. The essence of this method
appeared in many places in the literature
\cite{kicked-Ising-space-time-duality-1,dual-unitary-1,dual-kicked,dual-unitary--bernoulli,arul-perfect}, but its most
general formulation was 
given in \cite{claeys-lamacraft-emergent}. Here we explore this idea once more, extending it also to the cubic geometry.

A complex Hadamard matrix $H$ of size $N\times N$ has the following two properties: all its matrix elements are complex
numbers of modulus one, and the matrix is proportional to a unitary matrix. The review
\cite{complex-hadamard} discusses the history of research on Hadamard
matrices, their applications to combinatorial design theory and quantum information theory, and it also gives a
number of constructions and concrete examples for small sizes.

Let us also discuss the simplest examples for Hadamard matrices. First we note that
the Hadamard matrices form equivalence classes: two Hadamard matrices $A$ and $A'$ are in the same class if there
are diagonal unitary matrices $D_1$ and $D_2$ and permutation matrices $P_1$ and $P_2$ such that
\begin{equation}
  A'=P_2D_2AD_1P_1
\end{equation}
Generally it is enough to discuss examples up to such equivalence steps.

For $N=2$ it is known that the only equivalence class is given by the real Hadamard matrix
\begin{equation}
  \label{F2}
  A=
  \begin{pmatrix}
1 & 1 \\ 1& -1
  \end{pmatrix}
\end{equation}
For $N=3$ there is again only one equivalence class, which is given by
\begin{equation}
  \label{F3}
  A=
  \begin{pmatrix}
    1 & 1 & 1 \\
    1& \omega & \omega^2 \\
1 & \omega^2 & \omega \\
  \end{pmatrix},\qquad \omega=e^{2i\pi/3}.
\end{equation}
For a generic $N$ a symmetric complex Hadamard matrix is the Fourier matrix, given by
\begin{equation}
  \label{Fourier}
  F_{jk}=e^{\frac{2i\pi}{N}(j-1)(k-1)},\qquad j,k=1,\dots,N.
\end{equation}
The matrices \eqref{F2} and \eqref{F3} are special cases of the Fourier matrix at $N=2$ and $N=3$. 

\subsection{Dual unitary matrices}

In the case of square geometry, let us take four Hadamard matrices $A, B, C, E$ of size $N\times N$ and construct a new matrix $H$ of size $N^2\times
N^2$ (acting on the two-fold tensor product space) through the formula
\begin{equation}
  \label{fourfold}
  H_{ab}^{dc}=A_{a}^bB_{b}^cC_{c}^dE_{d}^a.
\end{equation}
There is no summation 
over $a,b,c,d$; this is just a product of the given matrix elements. Clearly we get $|H_{ab}^{cd}|=1$. The
geometrical interpretation is the following: if we regard the sites as the four vertices of a square, then the small
Hadamard matrices are attached to the four edges of the same square, and multiplication is taken 
component-wise.

It is easy to show that the matrix $\check U=H/N$ is unitary. To see this, we rewrite the matrix $\check U$ as
\begin{equation}
  \label{fourfold1a}
\check  U=\frac{1}{N}D^C  (E^t\otimes B)D^A,
\end{equation}
where $E$ and $B$ are the operators as given by the original Hadamard matrices, $^t$ denotes transpose, and $D^A$ and
$D^C$ are diagonal two-site operators given by
\begin{equation}
  \label{DADC}
  (D^A)_{ab}^{ab}=A_a^b,\qquad  (D^C)_{dc}^{dc}=C_c^d.
\end{equation}
It follows from the Hadamard properties that 
$D^A$ and $D^C$
are unitary, and supplemented with the unitarity of
$B/\sqrt{N}$ and $E^t/\sqrt{N}$ we see that $U$ is indeed unitary. Essentially the same proof was given in
\cite{claeys-lamacraft-emergent}. 

The construction \eqref{fourfold} is invariant with respect to the symmetries of the square: the 
transformations
lead to transpositions and permutations of the factors, but they do not change the structure of the formula. Therefore,
these $\check U$ matrices are also dual unitary.

Solutions which are invariant with respect to the geometrical symmetries are found when all matrices are equal and
symmetric. Thus we choose a single complex Hadamard matrix $A$ satisfying $A=A^t$ and we write
\begin{equation}
  \label{fourfold2}
  \check U_{ab}^{dc}=\frac{1}{N}A_{a}^bA_{b}^cA_{c}^dA_{d}^a.
\end{equation}
Each such $\check U$ is dual unitary and self dual.

In the case of $N=2$ we consider the Hadamard matrix
\begin{equation}
  \label{F2i}
  A'=
  \begin{pmatrix}
    1 & i \\ i & 1
  \end{pmatrix}.
\end{equation}
This is in the same equivalence class as $A$ in \eqref{F2}. 
It can be checked by direct computation that the gate of the kicked Ising model \eqref{kicked2} is obtained from
this matrix via \eqref{fourfold2} (up to an overall phase). 

For generic $N$ the quantum
circuits arising from the Fourier matrix were considered earlier in \cite{dual-kicked}.

\subsection{Cubic geometry}

\label{hadcubic}

The previous construction can be extended to the cubic geometry. 
Now we take 12 copies of a symmetric complex Hadamard matrix $A$ and ``attach'' them to the 12 edges of the cube. The
components 
of the operator $\check U$ are then obtained by taking the products of matrix elements of $A$.  An explicit formula for the spatially symmetric case is
written down as
\begin{equation}
  \check U= D^A_{14}D^A_{34}D^A_{23}D^A_{12}A_4A_3A_2A_1D^A_{14}D^A_{34}D^A_{23}D^A_{12},
\end{equation}
where $D^A$ is defined in \eqref{DADC}.
Each factor in this product represents one of the edges of the cube. The formula is a generalization of
\eqref{fourfold1a}. Unitarity is proven in a straightforward way, and the spatial symmetry follows from the construction.

The simplest case is that of the kicked Ising model $(N=2)$ in 2+1 dimensions, which was considered in
\cite{kicked-ising-3d}. In this case the chosen Hadamard matrix is again given by \eqref{F2i}, and 
the operator $\check U$ acting on 4 qubits can be written as 
\begin{equation}
  \check U=
  e^{i\frac{\pi}{4}(Z_1Z_2+Z_2Z_3+Z_3Z_4+Z_4Z_1)}
  e^{i\frac{\pi}{4}(X_1+X_2+X_3+X_4)}e^{i\frac{\pi}{4}(Z_1Z_2+Z_2Z_3+Z_3Z_4+Z_4Z_1)}.
\end{equation}
This operator can be written as $\check U=\frac{1}{2^{3/2}}H$, where $H$ is a complex Hadamard matrix of size $16\times
16$. This matrix is equivalent to a real Hadamard matrix, which can be seen from the equivalence between 
\eqref{F2} and \eqref{F2i}.

\section{Graph states}

\label{sec:graph}

Graph states are well known entanglement resources in quantum information theory. The main idea behind them is to take a
Hilbert space of $K$ sites, to start from a product state, and apply a set of commuting entangling operations which can
be encoded in a graph with $K$ vertices. In this work we apply the formalism and the main results of
\cite{AME-graph}. We assume that the local dimension $N$  is a prime 
number; extensions of the method to prime power dimensions
were treated in \cite{AME-graph-2}.

A graph state $\ket{\psi}$ on $K$ sites is characterized by a ``colored'' (or labeled) undirected graph with $K$
vertices. Each edge 
of the graph can take $N$ values (labels) given by $\{0,1,2,\dots,N-1\}$; these values are encoded in the incidence matrix
$I_{jk}$, where $j\ne k$ stand for two vertices. Self loops are disregarded.

The construction starts with a product state $\ket{\psi_0}$ which is prepared as
\begin{equation}
  \ket{\psi_0}=\otimes_{j=1}^K \ket{e}_j,\qquad \ket{e}=\frac{1}{\sqrt{N}}\left(\sum_{a=1}^N \ket{a}\right).
\end{equation}
We act on this product state with a sequence of two-site gates. We introduce the so-called controlled-$Z$ gate, which  is
a diagonal two-site operator  with matrix elements
\begin{equation}
  \ZZ_{ab}^{ab}=(\omega)^{a\cdot b},\qquad \omega=e^{2i\pi/N}.
\end{equation}
The matrix elements of $\ZZ$ are identical with those of the Fourier matrix
\eqref{Fourier} (up to an overall constant), but now the $N\times N$ matrix elements are arranged in a diagonal matrix of size
$N^2\times N^2$. In a certain sense the matrix $\ZZ$ is the dual of the Fourier matrix \eqref{Fourier}. The
transposition symmetry of 
the matrix $F$ translates into a space reflection symmetry of $\ZZ$.

Let us now draw a graph on the $K$ vertices with incidence matrix $I$. Then the corresponding graph state
is 
\begin{equation}
  \label{graphstate}
  \ket{\psi}=\prod_{1\le j<k\le K}  (\ZZ_{jk})^{I_{jk}}\ket{\psi_0}.
\end{equation}
Here $\ZZ_{jk}$ stands for the controlled-$Z$ operator acting on sites $j$ and $k$.
These operators are all diagonal in the given basis, therefore their product is well defined without specifying
the ordering. The resulting state is such that each component has equal magnitude, and entanglement arises from the
various phases that can appear.

The graph state is characterized by the incidence matrix $I$. States corresponding to two different graphs can be LU
equivalent; this is treated for example in \cite{graph-clifford-1,graph-clifford-2}. It is important that the formula
\eqref{graphstate} does not depend on $N$, but the vectors $\ket{\psi_0}$ and the operator $\ZZ$ are different for
different values of $N$. Therefore, the same graph can encode states with very different types of entanglement  as $N$
is varied. We also note that in certain cases the graph states are LU equivalent to ``classical solutions'' which are
treated in Section \ref{sec:classical}; for the treatment of these connections see for example \cite{four-AME,arul-perfect}. 

The entanglement properties of the states \eqref{graphstate} can be found from certain properties of the incidence
matrix. It was shown in \cite{AME-graph} that the state is maximally entangled with respect to a bipartition $S=A\cup
B$ if a certain reduced incidence matrix $I^{AB}$ has maximal rank. To be more concrete, let us take a graph with $K$
vertices, and divide the sets of vertices into two subsets $A$ and $B$ with $K/2$ sites. The reduced incidence matrix
$I^{AB}$ of size $K/2\times K/2$ is determined by the links connecting the different points in $A$ and $B$: if we order
the list of sites such that the first $K/2$ sites are in $A$ and the second $K/2$ sites are in $B$ then the original
incidence matrix is given in block form as
\begin{equation}
  I=
  \begin{pmatrix}
    I^{AA} & I^{AB} \\
    I^{BA} & I^{BB}
  \end{pmatrix},\qquad I^{BA}=(I^{AB})^t
\end{equation}
Maximal rank means that the rows (or columns) of $I^{AB}$ are linearly independent vectors over $\egesz_N$, or
alternatively, that the determinant of the matrix is non-zero in $\egesz_N$. In practice this is equivalent to the
condition
\begin{equation}
  \det I^{AB}\ne 0 \text{ mod } N
\end{equation}

It is very important that the condition is understood in $\egesz_N$: some graph states can be maximally entangled for
certain values of $N$, and less entangled for other values; examples are treated in \cite{AME-graph}.

It is easy to find a condition for the spatial invariance \eqref{psym}:
The graph must be invariant with respect to
the given geometric symmetry group. This requirement dramatically decreases the number of possibilities, but it also enables a
simple direct check of the maximal entanglement. Below we analyze the solutions in the different geometries.

In all our cases we choose a special convention for the reduced density matrix $I^{AB}$. We divide the set of $K$
vertices into two subsets $A=\{1,\dots,K/2\}$ and $B=\{K/2+1,\dots,K\}$, and we set the matrix element $(I^{AB})_{jk}$
with $j,k=1,\dots,K/2$ to be equal to the label of the edge connecting the sites with indices $j$ and $k+K/2$. We use the
ordering of the site indices explained in Subsection \ref{sec:opstate}.

\subsection{Dual unitary matrices}

Again we are dealing with the geometry of the square. Spatially invariant graphs on the square have just two free
parameters: the labels for the edges and for the diagonals of the square. Denoting them by $\alpha$ and $\beta$,
respectively,  we get
the reduced incidence matrix 
\begin{equation}
  I^{AB}=
  \begin{pmatrix}
     \beta &\alpha \\
    \alpha & \beta
  \end{pmatrix}.
\end{equation}
There is maximal entanglement if $\beta^2-\alpha^2\ne 0 \text{ mod } N$. This leads to the two conditions
\begin{equation}
  \beta+\alpha \ne 0 \text{ mod } N,\qquad \text{ and }   \beta-\alpha \ne 0 \text{ mod } N.
\end{equation}
For $N=2$ there are only two choices. 
The case of $\beta=0$ and $\alpha=1$ is a special case of the
construction \eqref{fourfold2} with Hadamard matrices. The case of $\alpha=0$ and $\beta=1$ is simply just the
identity along the diagonals dressed with one-site unitary operators. 

For $N=3$ the two conditions imply that either $\alpha=0$ or $\beta=0$, leading to essentially the same
configurations as for $N=2$.

New solutions appear for $N>3$. For example $\alpha=1$ and $\beta=2$ is a solution for every prime $N>3$.

Dual unitary matrices constructed from graph states were considered earlier in
\cite{dual-unitary--bernoulli,arul-perfect,sajat-dual-unitary}; in \cite{dual-unitary--bernoulli,arul-perfect} they were
called quantum cat maps. It is known that for $N=3$ there is a graph state, which is an AME, but it does not have
spatial invariance in the strict sense \cite{AME-graph}. We treat its classical version in Appendix \ref{app:luperm},
where we also show that it is spatially invariant in a weaker sense.

\subsection{Hexagonal geometry}

Now a graph state has three parameters. They correspond to the three types of vertices: those connecting nearest
neighbours, next-to-nearest neighbours, and sites on diagonals. Denoting the corresponding labels (in this order) as
$\alpha, \beta, \gamma$ we obtain the
reduced incidence matrix
\begin{equation}
  \label{ihexa}
  I^{AB}=
  \begin{pmatrix}
\gamma & \beta &     \alpha \\
\beta   &\gamma & \beta \\
    \alpha & \beta &\gamma \\
  \end{pmatrix}.
\end{equation}
Now we find the determinant
\begin{equation}
  \det I^{AB}=(\gamma-\alpha)(\gamma^2-2\beta^2+\gamma\alpha).
\end{equation}
Thus we get the two requirements
\begin{equation}\label{eq:2req}
  \alpha-\gamma\ne 0  \text{ mod } N,\text{ and } \gamma^2-2\beta^2+\gamma\alpha \ne 0  \text{ mod } N.
\end{equation}
For $N=2$ these requirements simplify to
\begin{equation}
  \alpha-\gamma\ne 0  \text{ mod } N,\text{ and } \gamma(\gamma+\alpha) \ne 0  \text{ mod } N.
\end{equation}
This dictates $\alpha=0$ and $\gamma=1$, and in this case $\beta$ can be chosen to be either 0 or 1. The resulting cases are:
\begin{itemize}
\item The solution with $\alpha=\beta=0$ and $\gamma=1$ is trivial, it corresponds to the identity along the diagonals
plus one-site unitary operators.
\item The solution $\alpha=0$ and $\beta=\gamma=1$ is non-trivial. More detailed
computations show that it is an AME. It was well known that  an AME with 6 qubits exists, and in fact a graph isomorphic
to the solution  $\alpha=0$ and $\beta=\gamma=1$ was presented in  \cite{AME-graph}, see Fig. 2.e in that work. However,
the hexagonal symmetry of that given graph was not pointed out there.
\end{itemize}

For $N\ge 3$ there are more solutions, and there are some solutions which work for every prime $N$. For example, the
choice $\alpha=1$, $\beta=1$ and $\gamma=0$ always satisfies (\ref{eq:2req}).

\subsection{Cubic geometry}

Now the graph consists of $K=8$ sites and there are 3 types of edges of the graph: those which correspond to the edges,
the face diagonals and the body diagonals of the cube. Denoting the corresponding labels as $\alpha$, $\beta$, $\gamma$
the reduced incidence matrix for a bipartition into two faces is written as
\begin{equation}
  I^{AB}=
  \begin{pmatrix}
    \gamma & \beta & \alpha  & \beta \\
 \beta    &  \gamma & \beta & \alpha \\
\alpha     & \beta  & \gamma & \beta \\
    \beta  & \alpha &  \beta & \gamma \\
  \end{pmatrix}.
\end{equation}
Now the determinant is
\begin{equation}
  \det I^{AB}=(\alpha-\gamma)^2(\alpha-2\beta+\gamma)(\alpha+2\beta+\gamma).
\end{equation}
For $N=2$ the determinant simplifies further to
\begin{equation}
  \det I^{AB}=(\alpha-\gamma)^4 \text{ mod } 2.
\end{equation}
Therefore the value of $\beta$ is irrelevant, and we get the only condition that $\alpha$ and
$\gamma$ need to be different. This leads to the following four solutions: 
\begin{itemize}
\item $\gamma=1$ and $\alpha=\beta=0$. This is the trivial solution, with non-interacting body diagonals.
\item $\alpha=1$ and $\beta=\gamma=0$. Now there are controlled $Z$-operators attached to the edges of the cube. This is
  the same state that is found via the Hadamard matrices in Section \ref{hadcubic}.
\item $\alpha=0$, $\beta=\gamma=1$ and  $\alpha=\beta=1$, $\gamma=0$. These are non-trivial new solutions.
\end{itemize}
Naturally, for $N\ge 3$ there are more solutions, which can be found easily by numerical inspection.

\subsection{Octahedral geometry}

Now there are $K=6$ sites and only two types of edges of the graph: those corresponding to the edges and the body
diagonals of the 
octahedron. Denoting their labels as $\alpha$ and $\gamma$, the reduced incidence matrix is
\begin{equation}
  \label{iocta}
  I^{AB}=
  \begin{pmatrix}
    \gamma &\alpha  & \alpha \\
\alpha    & \gamma & \alpha \\
  \alpha  & \alpha & \gamma
  \end{pmatrix}.
\end{equation}
We have
\begin{equation}
  \det I^{AB}=(\gamma-\alpha)^2(\gamma+2\alpha).
\end{equation}
For $N=2$ this simplifies to
\begin{equation}
  \det I^{AB}=(\gamma-\alpha)^2\gamma\text{ mod } 2.
\end{equation}
Therefore the only choice is $\gamma=1$, $\alpha=0$, which is the trivial solution with identity operations along the
diagonals. 

For $N\ge 3$ there are more solutions. Interestingly, the solution $\alpha=1$, $\gamma=0$ works for all prime
$N\ge 3$. In this case there is a controlled-$Z$ operator for each edge of the octahedron. 

We can see that the octahedral case is a specialization of the hexagonal case: the incidence matrix \eqref{iocta}
coincides with the special case $\alpha=\beta$ of \eqref{ihexa}.

\section{Classical solutions}

\label{sec:classical}

We call a unitary operator ``deterministic'' or ``classical'' if it is represented by a permutation matrix in a selected
basis. When the operator is applied successively, it gives rise to a deterministic walk over the elements of this
basis. In this section we consider multi-directional operators that have the classical property, i.e., they act as
permutations of basis elements in more than one direction. Instead of giving a general treatment right away we consider
the different geometries case by case.

We begin with the square geometry ($K = 4$). Now a unitary operator is represented by a matrix, whose matrix elements we
denote as $U_{a_1a_2}^{b_1b_2}$. The deterministic property combined with unitarity means that for all pairs of numbers
$(a_1,a_2)$ with  
$a_1,a_2\in \{1,\dots,N\}$ there is a unique pair $(b_1, b_2)$ with $b_1,b_2\in\{1,\dots,N\}$ such that $U_{a_1 a_2}^{b_1 b_2}=1$, and all other matrix elements are zero. Thus the classical unitary operator corresponds to a bijective mapping $X^2\to X^2$ with $X=\{1,\dots,N\}$, defined by the functions $b_1(a_1,a_2)$ and $b_2(a_1,a_2)$.
The operator corresponds to the quantum state given by
\begin{equation}
  \label{psymin}
  \ket{\psi}=\frac{1}{N}\sum_{a_1,a_2=1}^N \ket{a_1, a_2 ; b_1, b_2 },\qquad b_1 =b_1(a_1,a_2),\ b_2=b_2(a_1, a_2)\,.
\end{equation}
Such states have {\it minimal support}: It is not possible to satisfy the unitarity condition with states
with a fewer number of non-zero components.

The components of the state \eqref{psymin} define classical configurations of the square, where each vertex is assigned
a concrete value. We call the quartet of values $(a_1,a_2,b_1,b_2)$ an allowed configuration
in the state \eqref{psymin} if the corresponding basis element $\ket{a_1,a_2;b_1,b_2}$ appears in \eqref{psymin}.

Now the dual unitarity condition is the following: for each pair $(a_1,b_2)$ there is exactly one
pair $(a_2,b_1)$ such that the quartet  $(a_1,a_2,b_1,b_2)$ is an allowed configuration, and the same holds also vice
versa. In other words, the set of quartets defines a bijection between the pairs  $(a_1,b_2)$ and $(a_2,b_1)$. Combining
the unitarity requirements we obtain the following criterion: A deterministic operator is dual unitary, if the set of
the allowed configurations is such that a quartet $(a_1,a_2,b_1,b_2)$ can be uniquely identified by the pairs
$(a_1,a_2)$, or $(a_2,b_1)$, or $(b_1,b_2)$ or $(a_1,b_2)$. The choice of these pairs comes from all the possibilities of selecting
two neighbouring sites of the square.

Let us also consider the hexagonal geometry.
Now the unitary matrix given by components $U_{a_1a_2a_3}^{b_1b_2b_3}$ is such that for each triplet
of ``input variables''  $(a_1, a_2, a_3)$ there is a deterministic triplet of ``output variables'' $(b_1, b_2, b_3)$ such
that each $b_j$ is a function $b_j(a_1,a_2,a_3)$. Furthermore, the mapping between the triplets is bijective. 
In such a case
we construct the vector
\begin{equation}
   \label{psymin2}
  \ket{\psi}=\frac{1}{N^{3/2}}\sum_{a_1,a_2,a_3=1}^N \ket{a_1,a_2,a_3;b_1,b_2,b_3}.
\end{equation}
Now we call a sextet $(a_1,a_2,a_3,b_1,b_2,b_3)$ an allowed configuration in the state \eqref{psymin2} if the corresponding basis element $\ket{a_1,a_2,a_3;b_1,b_2,b_2}$ appears in \eqref{psymin2}. We associate the variables with the six
vertices of a hexagon, such that they are written down in an anti-clockwise manner (similar to the labels in Fig
\ref{fig:squarehexa}). 

Hexagonal unitarity requires that rotations of the state \eqref{psymin2} also correspond to unitary operators.  From this
we deduce a condition for the sextets: A state \eqref{psymin2} is hexagonal unitary 
if at any three neighbouring sites of the hexagon any of the $N^3$ possible triplets of values appears in exactly one allowed configuration.

For such states it is also customary to compile a table out of the allowed configurations.
For example, in the case of the dual unitary matrices the resulting table is of size $N^2\times 4$ and it includes the
quartets $(a_1,a_2,b_1,b_2)$. In the hexagonal cases the table includes the sextets $(a_1,a_2,a_3,b_1,b_2,b_3)$ and it
becomes of size $N^3\times 6$. More generally, if there are a total number of $K$ variables, then we obtain a table of
size $N^{K/2}\times K$.
These tables are generalizations of the {\it orthogonal arrays} known from combinatorial design theory. In the simplest
case an orthogonal array of strength $t$ with $N$ levels and $k$ factors is a table with $N^t$ rows, $k$ columns, filled
with numbers $1,\dots,N$
with the following ``orthogonality''
property: if one selects $t$ columns of the table in 
\emph{any possible way} then in the resulting
sub-table with $t$ columns and
$N^t$ rows, every one of the $N^t$ $t$-tuplets is present in exactly one 
row.

The tables constructed from the multi-directional unitary matrices satisfy
a relaxed version of the above ``orthogonality'' property:
if one selects $t$ columns of the table \emph{in certain allowed ways},
 then in the resulting sub-table with $t$ columns and
$N^t$ rows, every one of the $N^t$ $t$-tuplets is present in exactly one row. The allowed subsets of columns coincide
with the allowed 
subsystems in \eqref{bipartitions}. In the case of the planar arrangements these tables were already discussed in
\cite{planar-OA}, where they were called ``planar orthogonal arrays''. Our concept here is even more general because we
also allow higher dimensional arrangements, such as the cubic arrangement in three space dimensions.

In the most general case our state is of the form
\begin{equation}
 \label{psymin3}
  \ket{\psi}=\frac{1}{N^{K/2}}\sum_{a_1,\dots,a_{K/2}=1}^N \ket{a_1,\dots,a_{K/2};b_1,\dots,b_{K/2}},
\end{equation}
where it is understood that each $b_j$ is a function of the tuple $(a_1,\dots,a_{K/2})$. In such expressions the
ordering of the variables is such that $a_j$ and $b_j$ are sitting on antipodal sites
on the same diagonal, and the ordering of the
diagonals is the same as laid out in Section \ref{sec:opstate}.

\subsection{Spatially symmetric solutions}

Let us turn again to the spatially symmetric cases. Classical solutions with spatial symmetries can be found by studying
the orbits of classical configurations under the geometric symmetry group. Let us denote 
by $\ket{\phi}$ a classical configuration, that is
\begin{equation}
  \ket{\phi}=\ket{c_1,c_2,\dots,c_K},\qquad c_j\in \{1,\dots,N\}.
\end{equation}
We define its orbit as
\begin{equation}
  \label{orbit}
  \oo(\ket{\phi})=\sum_k \ket{\phi_k},
\end{equation}
where the summation is over all such (distinct) states $\ket{\phi_k}$, which can be obtained from $\ket{\phi}$ using a symmetry
transformation. More precisely, the summation includes states $\ket{\phi_k}$ for which there is a  $g\in G$ such that
$\PP_g\ket{\phi}=\ket{\phi_k}$. It is important 
that the summation is over the configurations, and not the symmetry transformations. This way each vector $\ket{\phi_k}$
is present in the summation only one time, with a coefficient of 1.

Let us consider examples for orbits. We take $N=2$ and the geometry of the square. Four examples for orbits are given by
\begin{equation}
  \label{orbits1}
  \begin{split}
  \oo(\ket{1111})&=\ket{1111},\\
   \oo(\ket{1212})&=\ket{1212}+\ket{2121},\\
    \oo(\ket{1122})&=\ket{1122}+\ket{1221}+\ket{2211}+\ket{2112},\\
       \oo(\ket{1112})&=\ket{1112}+\ket{1121}+\ket{1211}+\ket{2111}.
  \end{split}
\end{equation}
We encourage the reader to check that these are indeed the orbits, by using the ordering of the sites laid out in Section
\ref{sec:opstate}.

Clearly, spatially symmetric states $\ket{\psi}$ can be constructed as linear combinations of different orbits. Let us now take a
set of configurations $\{\ket{\phi_k}\}_{k=1,\dots,n}$ such that all of them belong to different orbits, and construct
the desired state as
\begin{equation}
  \label{psii}
 \ket{\psi}=\frac{1}{N^{K/2}} \sum_{k=1}^{n_o} \oo(\ket{\phi_k}).
\end{equation}
Here we did not specify the number $n_o$ of the different orbits that are needed (this will depend on the state), and we
included a normalization 
factor in anticipation of the final result.
A state defined as \eqref{psii} is manifestly symmetric, but it is not yet established that it describes a multi-directional unitary
operator. 

\subsection{Multi-directional unitarity}

Now we explore the unitarity conditions for the state \eqref{psii}.
It follows from
the spatial symmetry that unitarity needs to be checked with respect to only one of the allowed bipartitions. The
unitarity condition will be satisfied automatically for the other bipartitions. 

Returning to the  state \eqref{psymin3}, we see that it describes a unitary matrix if it describes a bijection between the tuplets
$(a_1,\dots,a_{K/2})$  and $(b_1,\dots,b_{K/2})$. This means that 
for every tuplet $(a_1,\dots,a_{K/2})$ there is precisely one
allowed configuration in the state which has the numbers $(a_1,\dots,a_{K/2})$ as the first $K/2$ numbers, and also that
for each tuplet $(b_1,\dots,b_{K/2})$ there is precisely one allowed configuration which has these numbers as the last
$K/2$ numbers. This condition is easy to check once an explicit form for \eqref{psymin3} is given. However, it is less
evident from the expression \eqref{psii}. 

In order to find and describe the solutions we need to introduce new definitions.

We say that an orbit \eqref{orbit} is non-overlapping if there are no two distinct configurations $\ket{\phi_1}$ and
$\ket{\phi_2}$ which both appear in the orbit and which have the same values for the first $K/2$ variables.
It is clear that a state $\ket{\psi}$ with the desired entanglement properties
can include  only non-overlapping orbits, because otherwise the classical unitarity would be broken already by this
single orbit.

For example, consider the geometry of the square, $N=2$, and the orbits given in eq. \eqref{orbits1}. It can be checked
that the first three orbits are 
non-overlapping, but the fourth orbit is overlapping. In that case the first two configurations have the same pair $(1,1)$
as the first two components.
Therefore, this fourth orbit cannot be used for our purposes.

We say that two orbits are mutually non-overlapping, if there are no two configurations $\ket{\phi_1}$ and
$\ket{\phi_2}$ from the first and the second orbit, respectively, such that $\ket{\phi_1}$ and
$\ket{\phi_2}$ have the same values for the first $K/2$ components.
States $\ket{\psi}$ with the desired properties can be obtained as
sums of orbits which are
non-overlapping and also mutually non-overlapping.

The last property that needs to be checked is that
 for each possible value of the tuplet $(a_1,a_2,\dots,a_{K/2})$
 there is an orbit containing an allowed configuration which includes these numbers as the first $K/2$ variables. We
 call this the criterion of completeness.

Having clarified these conditions it is straightforward to set up an algorithm to find the solutions.
For each geometry and each $N$ one needs to determine the list of orbits which are non-overlapping. Having determined
this list one needs to look for a selection of orbits such that each two selected orbits are mutually non-overlapping,
and the criterion of completeness is also satisfied. In simple cases these computations can be carried out by hand. In
fact we found certain solutions without the help of a computer. On the other hand, for bigger values of $K$ and $N$ it
is useful to write computer programs, so that all solutions can be found.

Let us discuss a particular solution as an example. We take the square geometry and $N=4$. We start with the orbit
\begin{equation}
  \begin{split}
    \oo(\ket{1234})=&\ket{1234}+\ket{2341}+\ket{3412}+\ket{4123}+\\
&    \ket{4321}+\ket{3214}+\ket{2143}+\ket{1432}.
  \end{split}
  \label{psi40}
\end{equation}
It can be seen that this orbit is non-overlapping. We select the following also non-overlapping orbits:
\begin{equation}
  \begin{split}
    \oo(\ket{aaaa})=&\ket{aaaa},\qquad a=1, 2, 3, 4\\
\oo(\ket{1313})=& \ket{1313}+\ket{3131}\\
\oo(\ket{2424})=& \ket{2424}+\ket{4242}.\\
\end{split}
\end{equation}
Once again, all six orbits are non-overlapping. Furthermore, we can observe that all of the seven orbits are mutually
non-overlapping. Finally, their sum is complete. This means that the state $\ket{\psi}$ given by
\begin{equation}
  \label{psi4}
  \begin{split}
      \ket{\psi}=&\frac{1}{4}\Big[\ket{1234}+\ket{2341}+\ket{3412}+\ket{4123}+\\
  &    \ket{4321}+\ket{3214}+\ket{2143}+\ket{1432}+\\
  &   \ket{1313}+\ket{3131}+ \ket{2424}+\ket{4242}+\\
&   \ket{1111}+\ket{2222}+ \ket{3333}+\ket{4444}\Big]
  \end{split}
\end{equation}
satisfies both the symmetry and the unitarity requirements.

The state above is specified by its seven orbits. However, six of these orbits consist of configurations which
correspond to identity operations along the diagonals. In order to simplify our presentation we introduce a separate
name for such cases. 
We say that a configuration is diagonally identical if it is of the form
\begin{equation}
  \ket{a_1,\dots,a_{K/2};b_1,\dots,b_{K/2}}=
  \ket{a_1,\dots,a_{K/2};a_1,\dots,a_{K/2}}.
\end{equation}
This means that two variables on the same diagonal always have the same value. The orbit of a diagonally identical
configuration consists of configurations of the same type. Such orbits are always non-overlapping. What is more, two
different orbits of the same type are always mutually non-overlapping.

Most of the solutions that we find will have a number of diagonally identical orbits. This was demonstrated in the
example \eqref{psi4}, which has 6 diagonally identical orbits, and only 1 orbit which is of different type. 

When describing the solutions we will always present the list of orbits which are not diagonally identical. Once such a
list is given, it is straightforward to complement the state with diagonally identical orbits. The algorithm for this is
the following:
We start with the sum over the orbits which are not diagonally identical.
For each tuplet $(a_1,\dots,a_{K/2})$ with $a_j=1,\dots,N$, $j=1,\dots,K/2$ we check whether there is an allowed
configuration in the 
sum of orbits
which has these values as the first $K/2$ variables. If there is no such configuration, then
we add the diagonally identical orbit of $(a_1,\dots,a_{K/2})$ to the state. Afterwards we proceed with checking the
tuplets, and eventually this leads to a complete solution.

Let us demonstrate this procedure on the example of the state \eqref{psi4}. Now the only orbit which is not diagonally
identical comes from the configuration $\ket{1234}$. This orbit gives the first 8 terms in \eqref{psi4}. Now we perform the
check over the pair of variables $(a_1,a_2)$. There are 16 such pairs, and we see that 8 of them are present in the
first orbit. The remaining ones are $(1,3)$, $(3,1)$, $(2,4)$, $(4,2)$ and the four pairs $(a,a)$ with
$a=1,\dots,4$. The resulting diagonally identical orbits are
\begin{equation}
  \begin{split}
    \oo(\ket{1313})&=\ket{1313}+\ket{3131}\\
      \oo(\ket{2424})&=\ket{2424}+\ket{4242}\\
  \end{split}
\end{equation}
and the four orbits with the single element $\ket{aaaa}$ with $a=1,\dots,4$. Adding these orbits we obtain the state
\eqref{psi4}.

This demonstrates that indeed it is enough to list the diagonally not-identical orbits. In order to further simplify the
notation, we will label an orbit as $[a_1a_2\dots b_1b_2\dots]$, and we will simply just list such
labels. In this notation the state \eqref{psi4} is ``encoded'' simply as $[1234]$. This means that the state is
constructed from the orbit $\oo(\ket{1234})$ and the completion as described above. For clarity we also treat another
example. Again for $K=4$ and $N=4$ consider the list of orbits
\begin{equation}
  \label{psi4a}
 [1424], [3344].  
\end{equation}
These two labels encode a solution as follows. The orbits are given by
\begin{equation}
  \begin{split}
    \oo(\ket{1424})&=\ket{1424}+\ket{2414}+\ket{4142}+\ket{4241}\\
    \oo(\ket{3344})&=\ket{3344}+\ket{3443}+\ket{4433}+\ket{4334}.
  \end{split}
\end{equation}
We complete these configurations as
\begin{equation}
  \label{psi4b}
  \begin{split}
  \ket{\psi}=&\frac{1}{4}\Big[\ket{1424}+\ket{2414}+\ket{4142}+\ket{4241}+\\
  &\ket{3344}+\ket{3443}+\ket{4433}+\ket{4334}+\\
  & \ket{1212}+\ket{2121}+\ket{1313}+\ket{3131}+\\
  &\ket{2323}+\ket{3232}+\ket{1111}+\ket{2222}\Big].
   \end{split}
\end{equation}
Now the first two lines come from the diagonally not-identical orbits given in the list \eqref{psi4a}, and the last two
lines come from the completion. Therefore, the short notation \eqref{psi4a} completely describes the final state
\eqref{psi4b}, which is a solution to all our requirements.

Other examples for the expansion of the list of orbits into the full state are presented in Appendix \ref{sec:appexp}.

\bigskip

We performed a search for all solutions using a computer program. We considered all geometries and small values of
$N$. 
In each case we increased $N$ until we found non-trivial solutions, which are not related to the identity matrix.

Before turning to concrete solutions we also discuss the local equivalences between the solutions.

\subsection{Equivalence classes}

 It was stated in
Section \ref{sec:spatially} that a natural LU equivalence between two spatially symmetric states is the one given by
\eqref{LU2}. However, in the classical case it is more natural to allow only classical equivalence steps. Therefore, we
say that two classical solutions $\ket{\psi_1}$ and $\ket{\psi_2}$ are ``weakly equivalent'' if there exists a one-site
permutation operator $V$, such that
\begin{equation}
  \label{LU3}
\ket{\psi}\quad\to\quad  (V\otimes V\otimes \dots\otimes V)\ket{\psi}
\end{equation}
Such a symmetric transformation preserves the structure of the orbits, it just re-labels the variables. 

Alternatively, we can also require stronger equivalence relations. One option is to allow the connection \eqref{LU2}
using arbitrary $SU(N)$ operators (not just permutations), but still requiring exact spatial symmetry. The other option is
to allow permutation operators, but to drop the spatial symmetry and allow a generic situation as in
\eqref{LU}. This gives
\begin{equation}
  \label{LUb}
  \ket{\psi}\quad\to\quad  (A^{(1)}\otimes A^{(2)}\otimes \dots\otimes A^{(K)})\ket{\psi},
  \end{equation}
where now each $A^{(j)}$ is a one-site permutation matrix. These stronger equivalence relations can connect classical
states which appear different if we consider the weak equivalence. On the other hand, they can also change the structure
of the orbits.

Consider for example the following dual unitary operator for $N=2$:
\begin{equation}
  \label{UXX}
  U=X\otimes X,
\end{equation}
where $X$ denotes the respective Pauli matrix. The operator \eqref{UXX} describes a spin-flip along the diagonals of the
square. Therefore it is trivial: there is no entanglement between the diagonals. Furthermore, it is classical: it is
represented by the state
\begin{equation}
  \ket{\psi_1}=\frac{1}{2}\oo(\ket{1122})=\frac{1}{2}\Big[
  \ket{1122}+\ket{1221}+\ket{2112}+\ket{2211}
  \Big].
\end{equation}
On the other hand, we can also consider the diagonally identical state corresponding to $U=1$:
\begin{equation}
  \ket{\psi_2}=\frac{1}{2}\Big[
  \ket{1111}+\ket{1212}+\ket{2121}+\ket{2222}
  \Big].
\end{equation}
The two states are not LU equivalent with respect to \eqref{LU3}: for $N=2$ the only one-site permutations are the
identity and the spin-flip, and their symmetric application leaves both states invariant. On the other hand, the strong
equivalence steps do connect them. Regarding the symmetric but quantum mechanical equivalence we can use the operator
\begin{equation}
 V=\frac{1-i}{2}
  \begin{pmatrix}
    1 & i \\ i & 1
  \end{pmatrix},\qquad V^2=X.
\end{equation}
Taking the ``square root'' of \eqref{UXX} we see that the equivalence \eqref{LU2}  holds with this choice for $V$.
Alternatively, we can also connect $\ket{\psi_1}$ and $\ket{\psi_2}$ using a non-symmetric permutation equivalence, for
example
\begin{equation}
  \ket{\psi_1}=(X\otimes X\otimes 1\otimes 1) \ket{\psi_2}.
\end{equation}
In this example the structure of the orbits is also changed. The state $\ket{\psi_1}$ is a single orbit,
whereas $\ket{\psi_2}$ consists of three diagonally identical orbits.

We define the weak and strong equivalence classes according to the equivalence steps discussed above.
The two equivalence classes are generally different, the strong equivalence classes can split into different weak classes.

Below we list the solutions up to the strong equivalence. This means that for each geometry and
each $N$ we give a list 
of states such that no two states can be transformed into each other via \eqref{LUb}.

For each equivalence class we choose a representative such that it has
the fewest orbits that are not diagonally identical within that
class. Returning to the examples of the states $\ket{\psi_1}$ and
$\ket{\psi_2}$ we would choose $\ket{\psi_2}$ because this state has
only diagonally identical orbits. In other words,
we choose representatives which are closest to the identity operation.

\subsection{Dual unitary matrices}
\label{sub:classdual}

For $N=2$  we found that there is only one equivalence class, namely the one given by the identity operator. Note
that using strong equivalence this class includes solutions such as \eqref{UXX}. 

For $N=3$ there are two equivalence classes, namely the one given by the identity operator, and another one given by $[1122]$ (which, unlike \eqref{UXX} in the case of $N=2$, is not equivalent to the identity).

A range of new solutions appear at $N=4$. These solutions are:
  \begin{enumerate}
    \item\, Identity; 
    \item\, [1234]; 
    \item\, [1122]; 
    \item\, [1424]; 
    \item\, [1424], [2343]; 
    \item\, [1424], [2343], [1122];   
    \item\, [1424], [2343], [1213]; 
    \item\, [1424], [2233]; 
    \item\, [1424], [3344]; 
    \item\, [1424], [3344], [1232]. 
  \end{enumerate}
  Apparently these solutions arise as sums of different orbits taken from a small list of orbits. Again, they each define an equivalence class with respect to the transformations \eqref{LUb}.

It is straightforward to generalize the computer programs to search for matrices with higher $N$. However, the runtime
of the program and the resulting list of solutions quickly becomes too long. Therefore we only publish results for small
$N$s, for which the list of equivalence classes is short enough to include in the article.

\subsection{Hexagonal geometry}
\label{sub:classhexa}

For $N=2$ there is only a single equivalence class, which is the class of the identity matrix.

New solutions appear at $N=3$. 
The full list of the solutions (strong equivalence classes) for $N=3$ is
\begin{enumerate}
  \item\, Identity; 
  \item\, [122 322]; 
  \item\, [122 322], [133 233]; 
  \item\, [122 322], [133 233], [211 311]; 
  \item\, [222 333], [232 323]; 
  \item\, [122 133], [123 132]; 
  \item\, [122 133], [232 323], [212 333], [113 223], [131 313]; 
  \item\, [111 232], [222 313], [333 121]; 
  \item\, [122 213], [232 323], [123 332], [132, 231];   
  \item\, [111 222],  [121 323], [212 333]. 
\end{enumerate}
As an example, Appendix \ref{sec:appexp} shows how to obtain the full state from the list of orbits in the case of
  solution No. 10.

\subsection{Cubic geometry}
\label{sub:classcubic}

The case of the cube is special: there are non-trivial solutions even in the case of $N=2$. There are two strong
equivalence classes, each having two solutions. One contains the identity and [1112~ 2221], [1122~ 2211], [1212~
2121]. The other contains [1112~ 2221] and [1122~ 2211], [1212~ 2121]. 

As an example, Appendix \ref{sec:appexp} shows how to obtain the full state for [1112~2221].

\subsection{Octahedral geometry}
\label{sub:classocta}
For $N=2$ there is just one strong equivalence class, namely the one given by the identity operator.

New solutions appear for $N=3$. The list of solutions (strong equivalence classes) is
\begin{enumerate}
  \item Identity;
  \item\, [222 333];
  \item\, [331 221]; 
  \item\, [331 332];
  \item\, [331 332], [221 223];
  \item\, [331 332], [221 223], [112 113].
  \end{enumerate}
As an example, Appendix \ref{sec:appexp} shows how to obtain the full state from the list of orbits in the case of
  solution No. 6.

Note that each one of these cases also appears in the list of the hexagonal solutions.
There are only 6 octahedral
solutions instead of the 10 hexagonal solutions: they are those cases which also satisfy the extra unitarity condition on top of those of the hexagonal case.
The list number of the matching hexagonal solutions is the following: 1: 1, 2: 5, 3: 6, 4: 2, 5: 3, 6: 4. (The list of orbits is different in some solutions because octahedral orbits are different from hexagonal ones.)

\section{Summary and outlook}

\label{sec:concl}

In this work we gave a unified treatment of multi-directional unitary operators and their corresponding multi-directional maximally entangled states.
Using this unified treatment, we have presented various ways to design such operators.
Most of 
these constructions had already
appeared in the literature before, but our treatment is unique: We centered the discussion around the geometric
properties of the states/operators, and this way we could treat all the different
constructions within the same
formalism.
Moreover, we believe that one of the geometries, namely the octahedral geometry has not been considered before.

We focused on spatially symmetric solutions
and in many cases we have found new concrete examples of multi-directional unitary operators. Our new results include
\begin{itemize}
\item The observation that the octahedral case is a specialization of the hexagonal case (with one more unitarity constraint).
\item Pointing out that there exists an AME of six qubits with exact hexagonal symmetry.
\item Finding graph states for the hexagonal, cubic and octahedral geometries.
\item Extending the construction of \cite{claeys-lamacraft-emergent} with Hadamard matrices to the cubic case.
\item Determining the classical solutions for small values of $N$ in all geometries.
\end{itemize}

There are various open problems which are worth pursuing.

First of all, it would be interesting to find a complete
description of the algebraic varieties at least in some simple cases, perhaps with additional symmetry properties. So
far the only complete description is for the four qubits in the square geometry (dual unitary matrices). It is likely
that complete solutions (explicit parametrizations) will not be found in other cases unless further restrictions are
imposed. 

It is also interesting to consider spatial invariance in a weak sense, as described at the end of Subsection \ref{sec:spatially} and
in the Appendix \ref{app:luperm}. Allowing a combination of permutations and LU operations opens up more possibilities
for such states.

Another interesting question is factorization in the cases with $K\ge 6$. Earlier works already
noted that multi-directional unitary operators can be constructed from products of the same objects with fewer
sites (tensor legs). For example, the work \cite{perfect-tangles} included a ``braiding'' construction for states in
the geometry of a perfect polygon with $K\ge 6$ sites, such that each braiding step involves a dual unitary
operator (this idea was originally suggested by Vaughan Jones). Similarly, gates for the cubic geometry were constructed
in \cite{ternary-unitary} with a product of a small number of 
dual unitary operators. Similar factorizations can be worked out in many cases, with small variations. Then it is also an
important question: Which final solutions can be factorized? In this work we did not touch this question, except for
pointing out that the operator in formula \eqref{D3Z} can be factorized in a special case. It is likely that most of our solutions
cannot be factorized, nevertheless this question deserves a proper study. Finding conditions for factorizability and/or
actually proving the impossibility of factorization in certain cases would help the general understanding of these objects.

In this work we focused only on the multi-directional unitary operators, and not on their applications. It would be
interesting to study the quantum circuits that arise from these objects. With the exception of the dual unitary
circuits, there has not been much progress yet. General statements about the behaviour of 
two-point functions were already given in \cite{triunitary,ternary-unitary}, but the dynamics of entanglement generation
has not yet been investigated. It would be interesting if certain models could be identified, where the entanglement
evolution could be computed, similar to some special dual unitary models such as the kicked Ising model.

\vspace{1cm}
{\bf Acknowledgments} 

We are thankful for discussions with
J\'anos Asb\'oth,
Wojciech Bruzda,
M\'at\'e Matolcsi,
Mil\'an Mosonyi,
P\'eter Vrana,
Mih\'aly Weiner,
Tianci Zhou,
Zolt\'an Zimbor\'as,
and Karol \.{Z}yczkowski.

\appendix

\section{Spatial invariance in a weak sense}

\label{app:luperm}

Here we provide two examples for a phenomenon discussed in Section \ref{sec:spatially}: states can have spatial
invariance in a weak sense, such that the spatial symmetry operations transform them into other vectors that are LU
equivalent to themselves.

The simplest example is in the case of the square geometry, which corresponds to the dual unitary operators. Now we
present the example on the level of the states. We consider four qubits, and two $SU(2)$-singlets which are prepared on
pairs of opposite sites, i.e. there is one singlet for each diagonal of the square. This gives (with some abuse of notation)
\begin{equation}
  \ket{\psi}=\left(\frac{\ket{12}-\ket{21}}{\sqrt{2}}\right)_{13}\left(\frac{\ket{12}-\ket{21}}{\sqrt{2}}\right)_{24}=
  \frac{1}{2}\left(
\ket{1122}-\ket{1221}-\ket{2112}+\ket{2211}
  \right).
\end{equation}
This state has maximal entanglement for the two bipartitions dictated by the geometry (namely $\{1,2\} \cup \{3,4\}$ and  $\{1,4\} \cup \{2,3\}$). Furthermore, it is invariant
with respect to reflections of the square.
However, a rotation of the square by $\pi/2$ results
in $-\ket{\psi}$, thus we obtain a ``projective representation'' of the invariance property.

Another example is given by a well known AME with $N=3$ and four parties. It is given by
\begin{equation}
  \label{AME3}
  \ket{\psi}=\sum_{a,b=0}^2 \ket{a,b,a+b,a-b}.
\end{equation}
Here the algebraic operations are performed in the finite field $\egesz_3$, and the indices for the basis vectors run
from 0 to 2. This state has maximal entanglement for all  bipartitions. And it is symmetric in a weak sense: any
permutation of its sites can be ``undone'' by local unitary transformations. For example, consider the permutation of
the first two sites. After a change of indices $a\leftrightarrow b$ we obtain the new representation
\begin{equation}
\PP_{12}  \ket{\psi}=\sum_{a,b=0}^2 \ket{a,b,a+b,-(a-b)}.
\end{equation}
We see that the only difference is in the last component. However, this can be transformed away by a unitary operator
acting on the last site,
which performs the change of basis elements $\ket{1}\leftrightarrow\ket{2}$, corresponding to negation
in $\egesz_3$. It 
can be checked that similar one-site transformations (or combinations thereof) exist for all rearrangements of the
sites. Therefore the state \eqref{AME3} is invariant in a weak sense.

\section{Explicit formulas for classical solutions}
\label{sec:appexp}

In Subsections \ref{sub:classdual}--\ref{sub:classocta} we give lists of classical multi-directional unitary operators arranged in various geometries.
These lists contain all the classical multi-directional unitary operators with certain low qudit dimensions $N$ that are completely symmetric with respect to their geometric arrangement.
The operators in these lists are written in a compact notation which enumerates only the diagonally non-identical orbits (for an explanation see the paragraphs following \eqref{psi4}).
In this Appendix we show examples of the expansion of this compact notation into the quantum state describing the operator. In the case of the square geometry (or dual unitary operators) an example is given in \eqref{psi4a}--\eqref{psi4b}. In this Appendix we consider the other three geometries. \\

\paragraph{Hexagonal geometry} Let us consider solution No. 10. in the list in Subsection \ref{sub:classhexa} with non-diagonal orbits [111 222], [121 323], [212 333]. The corresponding state is
\begin{equation}
\begin{aligned}
  \ket{\psi} =& \frac{1}{3\sqrt{3}}\bigg[
\big(\ket{111\,222} + 
\ket{112\,221} + 
\ket{122\,211} + 
\ket{222\,111} + 
\ket{221\,112} + 
\ket{211\,122}\big) \\
& +
\big(\ket{121\,323} + 
\ket{213\,231} + 
\ket{132\,312} + 
\ket{323\,121} + 
\ket{231\,213} + 
\ket{312\,132} \big) \\
& +
\big(\ket{212\,333} +
\ket{123\,332} + 
\ket{233\,321} + 
\ket{333\,212} + 
\ket{332\,123} + 
\ket{321\,233} 
\big) \\
& +
\big(\ket{113\,113} + 
\ket{131\,131} + 
\ket{311\,311}\big) + 
\big(\ket{313\,313} + 
\ket{133\,133} + 
\ket{331\,331}\big) \\
&+
\big(\ket{232\,232} + 
\ket{322\,322} + 
\ket{223\,223}\big) \bigg]
\end{aligned}
\end{equation}
In the r.h.s. distinct orbits are separated by parentheses.
The first three lines of the r.h.s. contain the three diagonally non-identical orbits. The last two lines contain diagonally identical orbits.

\paragraph{Cubic geometry}

  Let us consider one of the non-trivial solutions that appear in \ref{sub:classcubic} for $N=2$: [1112 2221]. The corresponding state is
\begin{equation}
\begin{aligned}
  \ket{\psi} =&
  \frac 14 \bigg[
  \big(\ket{1112\, 2221}+
  \ket{1121\, 2212}+
  \ket{1211\, 2122}+
  \ket{2111\, 1222} \\
 &+  \ket{2221\, 1112}+ 
  \ket{2212\, 1121}+
  \ket{2122\, 1211}+
  \ket{1222\, 2111}\big) \\
 &+ \big(\ket{1111\, 1111}\big)  
 +  \big(\ket{2222\, 2222}\big)
  +\big(\ket{1212\, 1212} 
  +\ket{2121\, 2121} \big)
 \\
 &+  \big(\ket{1122\, 1122} +
  \ket{1221\, 1221} +
  \ket{2211\, 2211} +
  \ket{2112\, 2112}
\big)\bigg]
\end{aligned}
\end{equation}
Again, distinct orbits are separated by parentheses. The first two lines contain the diagonally non-identical orbit. The rest contains identical orbits.

\paragraph{Octahedral geometry}
Let us consider solution No. 6. in the list in Subsection \ref{sub:classocta}. The diagonally non-identical orbits are [331 332], [221 223], [112 113].
\begin{equation}\begin{aligned}
    \ket{\psi} =& \frac{1}{3\sqrt 3}\bigg[ \big(\ket{331\,332} + 
\ket{332\,331} + 
\ket{313\,323} + 
\ket{323\,313} + 
\ket{133\,233} + 
\ket{233\,133} \big)  \\
&+
\big(\ket{221\,223} + 
\ket{223\,221} + 
\ket{212\,232} + 
\ket{232\,212} +
\ket{122\,322} + 
\ket{322\,122} \big) \\
&+
\big(\ket{112\,113} + 
\ket{113\,112} + 
\ket{121\,131} + 
\ket{131\,121} + 
\ket{211\,311} + 
\ket{311\,211} \big) \\
&+
\big(\ket{111\,111} \big)+ 
\big(\ket{222\,222} \big) +
\big(\ket{333\,333} \big) \\
&+
\big(\ket{123\,123}  +
\ket{231\,231} + 
\ket{312\,312}  
+ \ket{321\,321} +
\ket{213\,213} + \ket{132\, 132} \big)  \bigg]
  \end{aligned}
  \label{eq:hexa6}
\end{equation}
Distinct orbits are separated by parentheses. The first three lines contain the three diagonally non-identical orbits, the rest contains identity orbits.

Since the hexagonal constraints form a subset of the octahedral constraints, the state \eqref{eq:hexa6} also corresponds to the hexagonal unitary operator No. 4 in Subsection \ref{sub:classhexa}. Under the weaker hexagonal symmetry, the orbit in the last line, [123 123], splits into two orbits: [123~123] and [132 132].

\bigskip

\addcontentsline{toc}{section}{References}

\begin{thebibliography}{10}

\bibitem{dual-unitary-1}
B.~{Bertini}, P.~{Kos}, and T.~{Prosen}, ``{Exact Spectral Form Factor in a
  Minimal Model of Many-Body Quantum Chaos},''
  \href{http://dx.doi.org/10.1103/PhysRevLett.121.264101}{{\em Phys. Rev.
  Lett.} {\bf 121} (2018) no.~26, 264101},
  \href{http://arxiv.org/abs/1805.00931}{{\tt arXiv:1805.00931 [nlin.CD]}}.

\bibitem{dual-unitary-2}
B.~{Bertini}, P.~{Kos}, and T.~{Prosen}, ``{Entanglement Spreading in a Minimal
  Model of Maximal Many-Body Quantum Chaos},''
  \href{http://dx.doi.org/10.1103/PhysRevX.9.021033}{{\em Phys. Rev. X} {\bf 9}
  (2019) no.~2, 021033}, \href{http://arxiv.org/abs/1812.05090}{{\tt
  arXiv:1812.05090 [cond-mat.stat-mech]}}.

\bibitem{dual-unitary-3}
B.~Bertini, P.~Kos, and T.~Prosen, ``Exact Correlation Functions for
  Dual-Unitary Lattice Models in 1+1 Dimensions,''
  \href{http://dx.doi.org/10.1103/physrevlett.123.210601}{{\em Phys. Rev.
  Lett.} {\bf 123} (2019) no.~21, },
  \href{http://arxiv.org/abs/1904.02140}{{\tt arXiv:1904.02140
  [cond-mat.stat-mech]}}.

\bibitem{AME-review}
M.~{Enr{\'\i}quez}, I.~{Wintrowicz}, and K.~{{\.Z}yczkowski},
  \href{http://dx.doi.org/10.1088/1742-6596/698/1/012003}{``{Maximally
  Entangled Multipartite States: A Brief Survey},''} in {\em Journal of Physics
  Conference Series}, vol.~698 of {\em Journal of Physics Conference Series},
  p.~012003.
\newblock 2016.

\bibitem{jones-planar}
V.~F.~R. {Jones}, ``{Planar algebras, I},'' {\em arXiv Mathematics e-prints}
  (1999)  , \href{http://arxiv.org/abs/math/9909027}{{\tt arXiv:math/9909027
  [math.QA]}}.

\bibitem{kicked-Ising-space-time-duality-1}
M.~Akila, D.~Waltner, B.~Gutkin, and T.~Guhr, ``Particle-time duality in the
  kicked Ising spin chain,''
  \href{http://dx.doi.org/10.1088/1751-8113/49/37/375101}{{\em J. Phys. A} {\bf
  49} (2016) no.~37, 375101}.

\bibitem{sarang-lamacraft-kicked-DU}
S.~{Gopalakrishnan} and A.~{Lamacraft}, ``{Unitary circuits of finite depth and
  infinite width from quantum channels},''
  \href{http://dx.doi.org/10.1103/PhysRevB.100.064309}{{\em Phys. Rev. B} {\bf
  100} (2019) no.~6, 064309}, \href{http://arxiv.org/abs/1903.11611}{{\tt
  arXiv:1903.11611 [quant-ph]}}.

\bibitem{dual-unitary-4}
L.~{Piroli}, B.~{Bertini}, J.~I. {Cirac}, and T.~{Prosen}, ``{Exact dynamics in
  dual-unitary quantum circuits},''
  \href{http://dx.doi.org/10.1103/PhysRevB.101.094304}{{\em Phys. Rev. B} {\bf
  101} (2020) no.~9, 094304}, \href{http://arxiv.org/abs/1911.11175}{{\tt
  arXiv:1911.11175 [cond-mat.stat-mech]}}.

\bibitem{dual-gliders}
B.~{Bertini}, P.~{Kos}, and T.~{Prosen}, ``{Operator Entanglement in Local
  Quantum Circuits I: Chaotic Dual-Unitary Circuits},''
  \href{http://dx.doi.org/10.21468/SciPostPhys.8.4.067}{{\em SciPost Phys.}
  {\bf 8} (2020) no.~4, 067}, \href{http://arxiv.org/abs/1909.07407}{{\tt
  arXiv:1909.07407 [cond-mat.stat-mech]}}.

\bibitem{max-velocity}
P.~W. {Claeys} and A.~{Lamacraft}, ``{Maximum velocity quantum circuits},''
  \href{http://dx.doi.org/10.1103/PhysRevResearch.2.033032}{{\em Phys. Rev.
  Res.} {\bf 2} (2020) no.~3, 033032},
  \href{http://arxiv.org/abs/2003.01133}{{\tt arXiv:2003.01133 [quant-ph]}}.

\bibitem{tianci-max-velocity}
T.~{Zhou} and A.~W. {Harrow}, ``{Maximal entanglement velocity implies dual
  unitarity},'' {\em arXiv e-prints} (2022)  ,
  \href{http://arxiv.org/abs/2204.10341}{{\tt arXiv:2204.10341 [quant-ph]}}.

\bibitem{bruno-generic-entanglement}
A.~{Foligno} and B.~{Bertini}, ``{Growth of entanglement of generic states
  under dual-unitary dynamics},'' {\em arXiv e-prints} (2022)  ,
  \href{http://arxiv.org/abs/2208.00030}{{\tt arXiv:2208.00030
  [cond-mat.stat-mech]}}.

\bibitem{chaos-qchann}
P.~{Hosur}, X.-L. {Qi}, D.~A. {Roberts}, and B.~{Yoshida}, ``{Chaos in quantum
  channels},'' \href{http://dx.doi.org/10.1007/JHEP02(2016)004}{{\em JHEP} {\bf
  2016} (2016)  4}, \href{http://arxiv.org/abs/1511.04021}{{\tt
  arXiv:1511.04021 [hep-th]}}.

\bibitem{dual-unitary--bernoulli}
S.~{Aravinda}, S.~A. {Rather}, and A.~{Lakshminarayan}, ``{From dual-unitary to
  quantum Bernoulli circuits: Role of the entangling power in constructing a
  quantum ergodic hierarchy},''
  \href{http://dx.doi.org/10.1103/PhysRevResearch.3.043034}{{\em Phys. Rev.
  Res.} {\bf 3} (2021) no.~4, 043034},
  \href{http://arxiv.org/abs/2101.04580}{{\tt arXiv:2101.04580 [quant-ph]}}.

\bibitem{biunitary-permutations}
U.~Krishnan and V.~S. Sunder, ``On Biunitary Permutation Matrices and Some
  Subfactors of Index 9,'' {\em Transactions of the American Mathematical
  Society} {\bf 348} (1996) no.~12, 4691--4736.

\bibitem{biunitary-qinf1}
D.~J. {Reutter} and J.~{Vicary}, ``{Biunitary constructions in quantum
  information},'' {\em Higher Structures} {\bf 3} (2019) no.~1, 109--154,
  \href{http://arxiv.org/abs/1609.07775}{{\tt arXiv:1609.07775 [quant-ph]}}.

\bibitem{biunitary-qinf2}
V.~{Kodiyalam}, {Sruthymurali}, and V.~S. {Sunder}, ``{Planar algebras, quantum
  information theory and subfactors},'' {\em arXiv e-prints} (2019)  ,
  \href{http://arxiv.org/abs/1912.07228}{{\tt arXiv:1912.07228 [math.OA]}}.

\bibitem{triunitary}
C.~{Jonay}, V.~{Khemani}, and M.~{Ippoliti}, ``{Triunitary quantum circuits},''
  \href{http://dx.doi.org/10.1103/PhysRevResearch.3.043046}{{\em Phys. Rev.
  Res.} {\bf 3} (2021) no.~4, 043046},
  \href{http://arxiv.org/abs/2106.07686}{{\tt arXiv:2106.07686 [quant-ph]}}.

\bibitem{ternary-unitary}
R.~{Milbradt}, L.~{Scheller}, C.~{A{\ss}mus}, and C.~B. {Mendl}, ``{Ternary
  unitary quantum lattice models and circuits in $2 + 1$ dimensions},'' {\em
  arXiv e-prints} (2022)  , \href{http://arxiv.org/abs/2206.01499}{{\tt
  arXiv:2206.01499 [cond-mat.stat-mech]}}.

\bibitem{prosen-mikado}
Y.~{Kasim} and T.~{Prosen}, ``{Dual unitary circuits in random geometries},''
  {\em arXiv e-prints} (2022)  , \href{http://arxiv.org/abs/2206.09665}{{\tt
  arXiv:2206.09665 [cond-mat.stat-mech]}}.

\bibitem{AME-1}
P.~{Facchi}, G.~{Florio}, G.~{Parisi}, and S.~{Pascazio}, ``{Maximally
  multipartite entangled states},''
  \href{http://dx.doi.org/10.1103/PhysRevA.77.060304}{{\em Phys. Rev. A} {\bf
  77} (2008) no.~6, 060304}, \href{http://arxiv.org/abs/0710.2868}{{\tt
  arXiv:0710.2868 [quant-ph]}}.

\bibitem{AME-Helwig2}
W.~{Helwig}, W.~{Cui}, J.~I. {Latorre}, A.~{Riera}, and H.-K. {Lo}, ``{Absolute
  maximal entanglement and quantum secret sharing},''
  \href{http://dx.doi.org/10.1103/PhysRevA.86.052335}{{\em Phys. Rev. A} {\bf
  86} (2012) no.~5, 052335}, \href{http://arxiv.org/abs/1204.2289}{{\tt
  arXiv:1204.2289 [quant-ph]}}.

\bibitem{AME-Helwig3}
W.~{Helwig} and W.~{Cui}, ``{Absolutely Maximally Entangled States: Existence
  and Applications},'' {\em arXiv e-prints} (2013)  ,
  \href{http://arxiv.org/abs/1306.2536}{{\tt arXiv:1306.2536 [quant-ph]}}.

\bibitem{AMEcomb1}
D.~{Goyeneche} and K.~{{\.Z}yczkowski}, ``{Genuinely multipartite entangled
  states and orthogonal arrays},''
  \href{http://dx.doi.org/10.1103/PhysRevA.90.022316}{{\em Phys. Rev. A} {\bf
  90} (2014) no.~2, 022316}, \href{http://arxiv.org/abs/1404.3586}{{\tt
  arXiv:1404.3586 [quant-ph]}}.

\bibitem{four-AME}
M.~{Gaeta}, A.~{Klimov}, and J.~{Lawrence}, ``{Maximally entangled states of
  four nonbinary particles},''
  \href{http://dx.doi.org/10.1103/PhysRevA.91.012332}{{\em Phys. Rev. A} {\bf
  91} (2015) no.~1, 012332}, \href{http://arxiv.org/abs/1411.6178}{{\tt
  arXiv:1411.6178 [quant-ph]}}.

\bibitem{OA-book}
A.~S. Hedayat, N.~J. Sloane, and J.~Stufken, {\em Orthogonal Arrays: Theory and
  Applications}.
\newblock Springer, 1999.

\bibitem{AMEcomb2}
D.~{Goyeneche}, D.~{Alsina}, J.~I. {Latorre}, A.~{Riera}, and
  K.~{{\.Z}yczkowski}, ``{Absolutely maximally entangled states, combinatorial
  designs, and multiunitary matrices},''
  \href{http://dx.doi.org/10.1103/PhysRevA.92.032316}{{\em Phys. Rev. A} {\bf
  92} (2015) no.~3, 032316}, \href{http://arxiv.org/abs/1506.08857}{{\tt
  arXiv:1506.08857 [quant-ph]}}.

\bibitem{AMEcomb3}
D.~{Goyeneche}, Z.~{Raissi}, S.~{Di Martino}, and K.~{{\.Z}yczkowski},
  ``{Entanglement and quantum combinatorial designs},''
  \href{http://dx.doi.org/10.1103/PhysRevA.97.062326}{{\em Phys. Rev. A} {\bf
  97} (2018) no.~6, 062326}, \href{http://arxiv.org/abs/1708.05946}{{\tt
  arXiv:1708.05946 [quant-ph]}}.

\bibitem{euler36}
S.~A. {Rather}, A.~{Burchardt}, W.~{Bruzda}, G.~{Rajchel-Mieldzio{\'c}},
  A.~{Lakshminarayan}, and K.~{{\.Z}yczkowski}, ``{Thirty-six Entangled
  Officers of Euler: Quantum Solution to a Classically Impossible Problem},''
  \href{http://dx.doi.org/10.1103/PhysRevLett.128.080507}{{\em Phys. Rev.
  Lett.} {\bf 128} (2022) no.~8, 080507},
  \href{http://arxiv.org/abs/2104.05122}{{\tt arXiv:2104.05122 [quant-ph]}}.

\bibitem{euler36-explanation}
K.~{{\.Z}yczkowski}, W.~{Bruzda}, G.~{Rajchel-Mieldzio{\'c}}, A.~{Burchardt},
  S.~A. {Rather}, and A.~{Lakshminarayan}, ``{9 $\times$ 4 = 6 $\times$ 6:
  Understanding the quantum solution to the Euler's problem of 36 officers},''
  {\em arXiv e-prints} (2022)  , \href{http://arxiv.org/abs/2204.06800}{{\tt
  arXiv:2204.06800 [quant-ph]}}.

\bibitem{arul-perfect}
S.~A. {Rather}, S.~{Aravinda}, and A.~{Lakshminarayan}, ``{Construction and
  local equivalence of dual-unitary operators: from dynamical maps to quantum
  combinatorial designs},'' {\em arXiv e-prints} (2022)  ,
  \href{http://arxiv.org/abs/2205.08842}{{\tt arXiv:2205.08842 [quant-ph]}}.

\bibitem{AME-list}
F.~Huber and N.~Wyderka, ``Table of AME states.'' Online available, 2021.
\newblock \url{https://www.tp.nt.uni-siegen.de/+fhuber/ame.html}.

\bibitem{AME-bounds}
F.~{Huber}, C.~{Eltschka}, J.~{Siewert}, and O.~{G{\"u}hne}, ``{Bounds on
  absolutely maximally entangled states from shadow inequalities, and the
  quantum MacWilliams identity},''
  \href{http://dx.doi.org/10.1088/1751-8121/aaade5}{{\em J. Phys. A} {\bf 51}
  (2018) no.~17, 175301}, \href{http://arxiv.org/abs/1708.06298}{{\tt
  arXiv:1708.06298 [quant-ph]}}.

\bibitem{ads-code-1}
F.~{Pastawski}, B.~{Yoshida}, D.~{Harlow}, and J.~{Preskill}, ``{Holographic
  quantum error-correcting codes: toy models for the bulk/boundary
  correspondence},'' \href{http://dx.doi.org/10.1007/JHEP06(2015)149}{{\em
  JHEP} {\bf 2015} (2015)  149}, \href{http://arxiv.org/abs/1503.06237}{{\tt
  arXiv:1503.06237 [hep-th]}}.

\bibitem{holocode-review}
T.~{Kibe}, P.~{Mandayam}, and A.~{Mukhopadhyay}, ``{Holographic spacetime,
  black holes and quantum error correcting codes: A review},''
  \href{http://dx.doi.org/10.1140/epjc/s10052-022-10382-1}{{\em Eu. Phys. J. C}
  {\bf 82} (2022) no.~5, 463}, \href{http://arxiv.org/abs/2110.14669}{{\tt
  arXiv:2110.14669 [hep-th]}}.

\bibitem{perfect-tangles}
J.~{Berger} and T.~J. {Osborne}, ``{Perfect tangles},'' {\em arXiv e-prints}
  (2018)  , \href{http://arxiv.org/abs/1804.03199}{{\tt arXiv:1804.03199
  [quant-ph]}}.

\bibitem{planar-AME}
M.~{Doroudiani} and V.~{Karimipour}, ``{Planar maximally entangled states},''
  \href{http://dx.doi.org/10.1103/PhysRevA.102.012427}{{\em Phys. Rev. A} {\bf
  102} (2020) no.~1, 012427}, \href{http://arxiv.org/abs/2004.00906}{{\tt
  arXiv:2004.00906 [quant-ph]}}.

\bibitem{block-perfect-tensor}
R.~J. Harris, N.~A. McMahon, G.~K. Brennen, and T.~M. Stace,
  ``Calderbank-Shor-Steane holographic quantum error-correcting codes,''
  \href{http://dx.doi.org/10.1103/PhysRevA.98.052301}{{\em Phys. Rev. A} {\bf
  98} (2018)  052301}, \href{http://arxiv.org/abs/1806.06472}{{\tt
  arXiv:1806.06472 [quant-ph]}}.

\bibitem{Dicke-states-entanglement}
A.~{Burchardt}, J.~{Czartowski}, and K.~{{\.Z}yczkowski}, ``{Entanglement in
  highly symmetric multipartite quantum states},''
  \href{http://dx.doi.org/10.1103/PhysRevA.104.022426}{{\em Phys. Rev. A} {\bf
  104} (2021) no.~2, 022426}, \href{http://arxiv.org/abs/2105.12721}{{\tt
  arXiv:2105.12721 [quant-ph]}}.

\bibitem{Dicke-states-ent-2}
D.~W. {Lyons}, J.~R. {Arnold}, and A.~F. {Swogger}, ``{Local unitary classes of
  states invariant under permutation subgroups},''
  \href{http://dx.doi.org/10.1103/PhysRevA.105.032442}{{\em Phys. Rev. A} {\bf
  105} (2022) no.~3, 032442}, \href{http://arxiv.org/abs/2109.06921}{{\tt
  arXiv:2109.06921 [quant-ph]}}.

\bibitem{huse-dual-clifford}
G.~M. {Sommers}, D.~A. {Huse}, and M.~J. {Gullans}, ``{Crystalline Quantum
  Circuits},'' {\em arXiv e-prints} (2022)  ,
  \href{http://arxiv.org/abs/2210.10808}{{\tt arXiv:2210.10808 [quant-ph]}}.

\bibitem{sajat-dual-unitary}
M.~{Borsi} and B.~{Pozsgay}, ``Construction and the ergodicity properties of
  dual unitary quantum circuits,''
  \href{http://dx.doi.org/10.1103/PhysRevB.106.014302}{{\em Phys. Rev. B} {\bf
  106} (2022)  014302}, \href{http://arxiv.org/abs/2201.07768}{{\tt
  arXiv:2201.07768 [quant-ph]}}.

\bibitem{dual-ensembles-1}
S.~A. Rather, S.~Aravinda, and A.~Lakshminarayan, ``Creating ensembles of dual
  unitary and maximally entangling quantum evolutions,''
  \href{http://dx.doi.org/10.1103/physrevlett.125.070501}{{\em Phys. Rev.
  Lett.} {\bf 125} (2020) no.~7, }, \href{http://arxiv.org/abs/1912.12021}{{\tt
  arXiv:1912.12021 [quant-ph]}}.

\bibitem{bipartite-unitaries}
J.~Deschamps, I.~Nechita, and C.~Pellegrini, ``On some classes of bipartite
  unitary operators,''
  \href{http://dx.doi.org/10.1088/1751-8113/49/33/335301}{{\em J. Phys. A} {\bf
  49} (2016) no.~33, 335301}, \href{http://arxiv.org/abs/1509.06543}{{\tt
  arXiv:1509.06543 [quant-ph]}}.

\bibitem{planar-OA}
Y.-L. {Wang}, ``{Planar k-uniform states: a generalization of planar maximally
  entangled states},'' \href{http://dx.doi.org/10.1007/s11128-021-03204-y}{{\em
  Quant. Inf. Proc.} {\bf 20} (2021) no.~8, 271},
  \href{http://arxiv.org/abs/2106.12209}{{\tt arXiv:2106.12209 [quant-ph]}}.

\bibitem{invariant-perfect}
Y.~{Li}, M.~{Han}, M.~{Grassl}, and B.~{Zeng}, ``{Invariant perfect tensors},''
  \href{http://dx.doi.org/10.1088/1367-2630/aa7235}{{\em New J. Phys.} {\bf 19}
  (2017) no.~6, 063029}, \href{http://arxiv.org/abs/1612.04504}{{\tt
  arXiv:1612.04504 [quant-ph]}}.

\bibitem{diagonal-du-1}
S.~{Singh} and I.~{Nechita}, ``{Diagonal unitary and orthogonal symmetries in
  quantum theory},'' \href{http://dx.doi.org/10.22331/q-2021-08-09-519}{{\em
  Quantum} {\bf 5} (2021)  519}, \href{http://arxiv.org/abs/2010.07898}{{\tt
  arXiv:2010.07898 [quant-ph]}}.

\bibitem{diagonal-du-2}
S.~{Singh} and I.~{Nechita}, ``{Diagonal unitary and orthogonal symmetries in
  quantum theory: II. Evolution operators},''
  \href{http://dx.doi.org/10.1088/1751-8121/ac7017}{{\em J. Phys. A} {\bf 55}
  (2022) no.~25, 255302}, \href{http://arxiv.org/abs/2112.11123}{{\tt
  arXiv:2112.11123 [quant-ph]}}.

\bibitem{diagonal-du-3}
S.~{Singh}, N.~{Datta}, and I.~{Nechita}, ``{Ergodic theory of diagonal
  orthogonal covariant quantum channels},'' {\em arXiv e-prints} (2022)  ,
  \href{http://arxiv.org/abs/2206.01145}{{\tt arXiv:2206.01145 [quant-ph]}}.

\bibitem{kicked-ising-eredeti}
T.~{Prosen}, ``{Exact Time-Correlation Functions of Quantum Ising Chain in a
  Kicking Transversal Magnetic Field ---Spectral Analysis of the Adjoint
  Propagator in Heisenberg Picture---},''
  \href{http://dx.doi.org/10.1143/PTPS.139.191}{{\em Prog. Theor. Phys. Supp.}
  {\bf 139} (2000)  191--203}, \href{http://arxiv.org/abs/nlin/0009031}{{\tt
  arXiv:nlin/0009031 [nlin.CD]}}.

\bibitem{dual-kicked}
B.~{Gutkin}, P.~{Braun}, M.~{Akila}, D.~{Waltner}, and T.~{Guhr}, ``{Local
  correlations in dual-unitary kicked chains},''
  \href{http://dx.doi.org/10.1103/PhysRevB.102.174307}{{\em Phys. Rev. B} {\bf
  102} (2020)  174307}, \href{http://arxiv.org/abs/2001.01298}{{\tt
  arXiv:2001.01298 [cond-mat.stat-mech]}}.

\bibitem{dual-unit-param}
P.~W. Claeys and A.~Lamacraft, ``{Ergodic and non-ergodic dual-unitary quantum
  circuits with arbitrary local Hilbert space dimension},''
  \href{http://dx.doi.org/10.1103/physrevlett.126.100603}{{\em Phys. Rev.
  Lett.} {\bf 126} (2021) no.~10, },
  \href{http://arxiv.org/abs/2009.03791}{{\tt arXiv:2009.03791 [quant-ph]}}.

\bibitem{claeys-lamacraft-emergent}
P.~W. {Claeys} and A.~{Lamacraft}, ``Emergent quantum state designs and
  biunitarity in dual-unitary circuit dynamics,''
  \href{http://dx.doi.org/10.22331/q-2022-06-15-738}{{\em Quantum} {\bf 6}
  (2022)  738}, \href{http://arxiv.org/abs/2202.12306}{{\tt arXiv:2202.12306
  [quant-ph]}}.

\bibitem{complex-hadamard}
W.~{Tadej} and K.~{Zyczkowski}, ``{A concise guide to complex Hadamard
  matrices},'' \href{http://dx.doi.org/10.1007/s11080-006-8220-2}{{\em Open.
  Syst. Inf. Dyn.} {\bf 13} (2006)  133},
  \href{http://arxiv.org/abs/quant-ph/0512154}{{\tt arXiv:quant-ph/0512154
  [quant-ph]}}.

\bibitem{kicked-ising-3d}
T.-C. {Lu} and T.~{Grover}, ``{Spacetime duality between localization
  transitions and measurement-induced transitions},''
  \href{http://dx.doi.org/10.1103/PRXQuantum.2.040319}{{\em PRX Quantum} {\bf
  2} (2021) no.~4, 040319}, \href{http://arxiv.org/abs/2103.06356}{{\tt
  arXiv:2103.06356 [quant-ph]}}.

\bibitem{AME-graph}
W.~Helwig, ``Absolutely Maximally Entangled Qudit Graph States,''
  \href{http://arxiv.org/abs/1306.2879}{{\tt arXiv:1306.2879 [quant-ph]}}.

\bibitem{AME-graph-2}
L.~Chen and D.~L. Zhou, ``Graph states of prime-power dimension from
  generalized CNOT quantum circuit,''
  \href{http://dx.doi.org/10.1038/srep27135}{{\em Sci. Rep.} {\bf 6} (2016)
  27135}, \href{http://arxiv.org/abs/1507.05386}{{\tt arXiv:1507.05386
  [quant-ph]}}.

\bibitem{graph-clifford-1}
M.~{van den Nest}, J.~{Dehaene}, and B.~{de Moor}, ``{Graphical description of
  the action of local Clifford transformations on graph states},''
  \href{http://dx.doi.org/10.1103/PhysRevA.69.022316}{{\em Phys. Rev. A} {\bf
  69} (2004) no.~2, 022316}, \href{http://arxiv.org/abs/quant-ph/0308151}{{\tt
  arXiv:quant-ph/0308151 [quant-ph]}}.

\bibitem{graph-clifford-2}
M.~{Bahramgiri} and S.~{Beigi}, ``{Graph States Under the Action of Local
  Clifford Group in Non-Binary Case},'' {\em arXiv e-prints} (2006)  ,
  \href{http://arxiv.org/abs/quant-ph/0610267}{{\tt arXiv:quant-ph/0610267
  [quant-ph]}}.

\end{thebibliography}

\providecommand{\href}[2]{#2}\begingroup\raggedright\endgroup

\end{document}